# Spin-free exact two-component linear response coupled cluster theory for estimation of frequency-dependent second-order property

Sudipta Chakraborty[a], Tamoghna Mukhopadhyay[a] and Achintya Kumar Dutta*,[a]

[a]*Department of Chemistry, Indian Institute of Technology Bombay, Mumbai-400076, India*

## Abstract

We have presented the theory, implementation, and benchmark results for the one-electronic variant of spin-free exact two-component (SFX2C1e) linear response coupled cluster (LRCCSD) theory for static and dynamic polarizabilities of atoms and molecules in the spin-adapted formulation. The resolution of identity (RI) approximation for two-electron integrals has been used to reduce the computational cost of the calculation and has been shown to have a negligible effect on accuracy. The calculated static and dynamic polarizability values agree very well with the more expensive X2C-LRCCSD and experimental results. Our calculated results show that accurate predictions of polarizabilities of atoms and molecules containing heavy atoms require the use of a large basis set containing an adequate number of diffuse functions, in addition to accounting for electron correlation and relativistic effects.

**achintya@chem.iitb.ac.in*

## 1. Introduction

The atomic and molecular response properties in the presence of external fields (electric or magnetic) stand as a cornerstone for studying linear and nonlinear optical phenomena[1]. Its significance spans a wide spectrum of applications, impacting fields from drug[2] design to the optimal functioning of optoelectronic devices, e.g. piezoelectric materials.[3,4] Quantum mechanical calculations of polarizabilities involve determining induced polarization as a function of the applied electric field. This is achieved through a time-independent or time-dependent perturbation calculation that accounts for the interaction between light and matter. In the last few decades, significant progress has been reported in *ab initio* calculation of response properties calculation[5–16]. However, most of them are formulated in a non-relativistic framework. In heavy elements, the influence of relativistic effects on response properties becomes increasingly pronounced. These relativistic effects can dramatically affect the electronic structure, leading to deviations from the predictions obtained from non-relativistic quantum mechanical calculations.[17] The incorporation of relativistic effects in wavefunction based calculations at the mean-field level is performed through the Dirac-Hartree-Fock (DHF)[18] method, which generally relies on a four-component Dirac Coulomb (DC) Hamiltonian. The static and dynamic polarizability calculations using coupled perturbed DHF have been reported in the literature[19]. However, electron correlation effects have a significant contribution to the accuracy of the polarizability values[20] for the non-relativistic case. Four-component Density Functional Theory (4c-DFT)[21] offers a computationally efficient way to incorporate correlation effects into the relativistic polarizability calculation. However, similar to its non-relativistic analogue, the accuracy of the relativistic DFT results is strongly dependent upon the used exchange-correlation functional and there is no systematic way to improve it[22,23]. Wave function based electron correlation methods lead to higher accuracies in calculated polarizability values.[12,24–26] Among the various wavefunction-based electron correlation methods, the coupled cluster method is considered to be highly accurate and systematically improvable[27]. Recently, Gomes and co-workers have implemented a linear response method based on a two-component relativistic Hamiltonian for polarizability and other second-order properties[28].

The coupled cluster[29,30] method is known for its high computational cost and scales as iterative $\mathcal{O}(N^6)$ power of the basis set even at the single and doubled truncation of the cluster operator (CCSD). The use of a relativistic Hamiltonian leads to a further increase in the computational cost of the coupled cluster calculation, and a relativistic CCSD[31–33] method based on the two or four-component Hamiltonian is at least 32 times costlier than its non-relativistic analogue.[26] The main reason behind elevation in the computational cost is the increase in the dimension due to four-component nature of the wave function comprising of large and small component spinors. One needs to uncontract the basis set to generate the necessary basis function for the small component using the kinetic balance condition[34,35]. At the same time, one must store complex numbers for the Hamiltonian matrix elements, which also leads to increased computational cost.

To simplify this four-component formalism, two-component (quasi-relativistic) approaches[36–38] have developed. This can be achieved by a unitary transformation known as the Foldy-Wouthuysen (FW) transformation[39]. However, except for free electrons, performing this transformation needs some approximations. An alternative approach, known as the Normalized Elimination of the Small Component (NESC) formalism[40–43], is a computationally straightforward method that allows for exact solutions of the Dirac equation for fermionic systems. The primary reason behind this rise in computational cost in a two or four component calculation is spin-orbit coupling (SOC), which disrupts the spin-symmetry, significantly increasing the number of molecular orbital integrals (MO)[44]. Although SOC effects are important in various scenarios, such as calculating energy differences in electronically degenerate states or simulating SOC-induced spin-forbidden transitions, they often have a relatively minor impact on many other properties[45]. In such cases, a scalar-relativistic approach that omits SOC effects may be sufficient, as the SOC corrections are reflected in the second-order of the perturbative expansion for closed-shell systems calculations[17]. Dyall's method of decoupling scalar-relativistic (spin-free) and SOC (spin-dependent) effects has led to the development of spin-free four-component Dirac-Coulomb (SFDC) approaches[46]. One can get a further simplification by using the exact two component (X2C)[37,47,48] approach within the spin-free framework. The X2C method comes in various flavors. In its simplest form, it diagonalizes the one-electron DC Hamiltonian matrix and uses the resulting two-component Hamiltonian matrix alongside the untransformed two-electron interactions. This method is referred to as X2C1e[49], and the corresponding spin-free version is called the SFX2C1e method. It has very little additional cost over non-relativistic calculations. The SFX2C1e coupled cluster method has been implemented for ground state energy, gradient[44,49,50], harmonic frequencies[51], and excited states[52]. In this work, we have extended the SFX2C1e coupled cluster method to calculate static and dynamic dipole polarizabilities using the coupled cluster linear response framework.

## 2. Theory

### *2.1. Spin separation of relativistic Hamiltonian*

In presence of an external electric field or potential $V$, the Dirac equation for an electron can be represented in two-spinor form as[53,54]

$$(V-E)\psi^L + c.(\vec{\sigma}.\vec{p})\psi^S = 0 \tag{1}$$

$$c(\vec{\sigma}.\vec{p})\psi^L + (V-E-2mc^2)\psi^S = 0 \tag{2}$$

Now in matrix formulation, the Dirac Hamiltonian ($\mathbf{H}^D$) and wavefunction ($\Psi$) should take the following form:

$$\mathbf{H}^D = \begin{pmatrix} V & c\vec{\sigma}.\vec{p} \\ c\vec{\sigma}.\vec{p} & V-2c^2 \end{pmatrix}, \qquad \Psi = \begin{pmatrix} \psi^L \\ \psi^S \end{pmatrix} \tag{3}$$

where $\vec{\sigma}$ and $\vec{p}$ represent Pauli's spin matrices and momentum vector, respectively and $\Psi$ consists of large($\psi^L$) and small components ($\psi^S$) of the four-component wave function. The structure of $\mathbf{H}^D$ matrix reveals that the diagonal block is spin-independent, and the off-diagonal block is spin-dependent. So, one can separate the spin-free and spin-dependent parts by constructing two separate terms from the partitioning of diagonal and off-diagonal blocks. Moreover, the small component is also represented in terms of a pseudo-large component $\phi^L$ based on kinetic balance approximation[53,54] as

$$\left|\psi^S\right\rangle = \frac{(\vec{\sigma}.\vec{p})}{2c}\left|\phi^L\right\rangle \tag{4}$$

where both small and large components are expanded on the same basis as

$$\left|\psi_p^L\right\rangle = c_{\mu p}^L\left|\chi_\mu\right\rangle \tag{5}$$

$$\left|\psi_p^S\right\rangle = c_{\mu p}^S\frac{\vec{\sigma}\cdot\vec{p}}{2c}\left|\chi_\mu\right\rangle \tag{6}$$

Here, $\left|\chi_\mu\right\rangle$ is the atomic basis function and $c_{\mu p}^L$, $c_{\mu p}^S$ are the corresponding coefficients for large and small components, respectively. One can achieve the spin separation from the one-electron Dirac Hamiltonian by writing the modified Dirac Hamiltonian $\mathbf{H}^M$ matrix as

$$\mathbf{H}^{M}\begin{pmatrix}\mathbf{C}^{\mathrm{L}}\\ \mathbf{C}^{\mathrm{S}}\end{pmatrix}=\begin{pmatrix}\mathbf{S} & \mathbf{0}\\ \mathbf{0} & \frac{1}{2c^{2}}\mathbf{T}\end{pmatrix}\begin{pmatrix}\mathbf{C}^{\mathrm{L}}\\ \mathbf{C}^{\mathrm{S}}\end{pmatrix}E \tag{7}$$

where,

$$\mathbf{H}^{\boldsymbol{M}}=\begin{pmatrix}\mathbf{V} & \mathbf{T}\\ \mathbf{T} & \frac{1}{4\boldsymbol{c}^{2}}\mathbf{W}-\mathbf{T}\end{pmatrix} \tag{8}$$

and $\mathbf{S},\mathbf{T},\mathbf{V}$ and $\mathbf{W}$ matrix elements are expressed as

$$[\mathbf{S}]_{\boldsymbol{\mu\nu}}=\langle\chi_{\mu}\,|\,\chi_{\nu}\rangle \tag{9}$$

$$[\mathbf{T}]_{\boldsymbol{\mu\nu}}=\left\langle\chi_{\mu}\left|\frac{\vec{p}^{2}}{2}\right|\chi_{\nu}\right\rangle \tag{10}$$

$$[\mathbf{V}]_{\boldsymbol{\mu\nu}}=\langle\chi_{\mu}\,|\hat{V}\,|\,\chi_{\nu}\rangle \tag{11}$$

$$[\mathbf{W}]_{\boldsymbol{\mu\nu}}=\langle\chi_{\mu}\,|\,(\vec{\sigma}\cdot\vec{p})\hat{V}(\vec{\sigma}\cdot\vec{p})\,|\,\chi_{\nu}\rangle \tag{12}$$

The modified Dirac Hamiltonian still consists of the spin-dependent term due to the presence of $\mathbf{W}$. Using Dirac identity, we can separate the matrix elements of $\mathbf{W}$ into spin-free $[\mathbf{W}]_{\boldsymbol{\mu\nu}}{}^{SF}$ and spin-dependent $[\mathbf{W}]_{\boldsymbol{\mu\nu}}{}^{SD}$ parts

$$[\mathbf{W}]_{\boldsymbol{\mu\nu}}=[\mathbf{W}]_{\boldsymbol{\mu\nu}}{}^{SF}+[\mathbf{W}]_{\boldsymbol{\mu\nu}}{}^{SD} \tag{13}$$

where,

$$[\mathbf{W}]_{\mu\nu}^{SF}=\langle\chi_{\mu}\left|\left(\vec{p}.V.\vec{p}\right)\right|\chi_{\mu}\rangle,\ \ [\mathbf{W}]_{\mu\nu}^{SD}=\langle\chi_{\mu}\left|\left(i\sigma.\left(\vec{p}\times V\vec{p}\right)\right)\right|\chi_{\mu}\rangle \tag{14}$$

and now the modified wave function is

$$\psi^{M}=\begin{pmatrix}\psi^{L}\\ \phi^{L}\end{pmatrix} \tag{15}$$

So, the explicit definition of modified spin-separated Dirac Hamiltonian is

$$\mathbf{H}^{M}=\begin{pmatrix}\mathbf{V} & \mathbf{T}\\ \mathbf{T} & [\mathbf{W}^{SF}-\mathbf{T}]/4c^{2}\end{pmatrix}+\begin{pmatrix}\mathbf{0} & \mathbf{0}\\ \mathbf{0} & \mathbf{W}^{SD}/4c^{2}\end{pmatrix}=\mathbf{H}^{SF}+\mathbf{H}^{SD} \tag{16}$$

The solution of this modified Hamiltonian exactly matches the solution of the original one electron Dirac Hamiltonian, which signifies that the method used to separate out the spin-dependent and spin-independent part consists of no approximation[17].

The computational overhead due to the presence of the small component can be reduced by removing the small component from the wave function, which is done by Foldy-Wouthuysen (FW) unitary transformation.

$$\mathbf{U}^{\dagger}\mathbf{H}^{M}\mathbf{U} = \begin{pmatrix} \mathbf{H}_{+}^{FW} & 0 \\ 0 & \mathbf{H}_{-}^{FW} \end{pmatrix} \tag{17}$$

where,

$$\mathbf{U} = \begin{pmatrix} 1 & -\mathbf{X}^{\dagger} \\ \mathbf{X} & 1 \end{pmatrix} \begin{pmatrix} \mathbf{R} & 0 \\ 0 & \mathbf{R}^{/} \end{pmatrix} = \mathbf{U}^{X}\mathbf{U}^{R} \tag{18}$$

In this transformation $\mathbf{U}^{X}$ takes care of the block diagonalization and $\mathbf{U}^{R}$ does the normalization on the modified Hamiltonian $\mathbf{H}^{M}$ . The operator $\mathbf{X}$ connects the small and large component coefficients.

$$\mathbf{X}C^{L} = C^{S} \tag{19}$$

The coefficient of the two-component wave function after the transformation is related to the large component one by the operator $\hat{\mathbf{R}}$

$$C_{FW}^{2C} = \frac{C^{L}}{\hat{R}} \tag{20}$$

where R is expressed as;

$$\hat{\mathbf{R}} = \mathbf{S}^{-\frac{1}{2}} \left( \mathbf{S}^{-\frac{1}{2}} \tilde{\mathbf{S}} \mathbf{S}^{-\frac{1}{2}} \right)^{-\frac{1}{2}} \mathbf{S}^{\frac{1}{2}} \tag{21}$$

and

$$\tilde{\mathbf{S}} = \mathbf{S} + \frac{1}{2c^{2}} \mathbf{X}^{\dagger}\mathbf{T}\mathbf{X} \tag{22}$$

The solution of $\mathbf{H}_{+}^{FW}$ will correspond to the orbital energies, and through this FW transformation will result:

$$\mathbf{H}_{+}^{FW} C_{FW}^{2C} = \mathbf{E}_{+}\mathbf{S}C_{FW}^{2C} \tag{23}$$

where $\mathbf{E}_{+}$ is the matrix corresponds to orbital energies and the $\mathbf{H}_{+}^{FW}$ is expressed as

$$\mathbf{H}_{+}^{FW} = \hat{\mathbf{R}}^{\dagger}\mathbf{L}^{NESC}\hat{\mathbf{R}} \tag{24}$$

with

$$\mathbf{L}^{NESC} = \mathbf{V} + \mathbf{T}\mathbf{X} + \mathbf{X}^{\dagger}\mathbf{T} - \mathbf{X}^{\dagger}\mathbf{T}\mathbf{X} + \frac{1}{4c^{2}} \mathbf{X}^{\dagger}\mathbf{W}\mathbf{X} \tag{25}$$

The spin-free X2C one electron (SFX2C1e) method is obtained by taking the spin-free part $\mathbf{H}^{SF}$ of the modified Hamiltonian $\mathbf{H}^{M}$ and subsequently performing the FW transformation. One can simply use this two-component FW transformed spin-free modified Hamiltonian to replace the one electron part of the non-relativistic Hamiltonian and use the non-relativistic two-electron atomic orbital integrals without any transformation. This will help to incorporate the scalar relativistic effect in SCF and electron-correlation calculations with negligible additional computational cost over a non-relativistic calculation.

### *2.2. Linear response coupled cluster method*

#### *(a) Time-independent coupled cluster theory*

In the coupled cluster method, the correlated ground state wave function can be generated by the action of an exponential ansatz over the reference (e.g., Hartree-Fock) wave function $\left|\Phi_0\right\rangle$ as

$$\left|\psi_{\mathrm{CC}}\right\rangle = e^{\hat{T}}\left|\Phi_0\right\rangle \tag{26}$$

The $\hat{T}$ is the cluster operator, which has the form of

$$\hat{T} = \hat{T}_1 + \hat{T}_2 + \hat{T}_3 + .... + \hat{T}_N = \sum_{p}^{N} \hat{T}_p \tag{27}$$

where

$$\hat{T}_p = \sum_{\substack{i>j>k>... \\ a>b>c>...}} t_{ijk...}^{abc...}\{\hat{a}_a^{\dagger}\hat{a}_b^{\dagger}\hat{a}_c^{\dagger}...\hat{a}_k\hat{a}_j\hat{a}_i...\} \tag{28}$$

One can write the cluster operator in a compact format as

$$\hat{T} = \sum_{\mu} t_{\mu}\hat{\tau}_{\mu}\,|\Phi_0\rangle \tag{29}$$

where $\hat{\tau}_{\mu}$ is used as a shorthand notation for the series of second quantized operators as,

$$\hat{\tau}_{\mu} = \{\hat{a}_a^{\dagger}\hat{a}_b^{\dagger}\hat{a}_c^{\dagger}...\hat{a}_k\hat{a}_j\hat{a}_i...\} \tag{30}$$

and $t_{\mu}$ is the corresponding amplitude.

One can use the generic notation $|\,\mu\rangle$ for excited determinants depending on the level of excitation.

$$|\,\mu\rangle = \left|\Phi_i^a\right\rangle, \left|\Phi_{ij}^{ab}\right\rangle, \left|\Phi_{ijk...}^{abc...}\right\rangle \tag{31}$$

Where $\left|\Phi_i^a\right\rangle, \left|\Phi_{ij}^{ab}\right\rangle, \left|\Phi_{ijk...}^{abc...}\right\rangle$ are the singly, doubly and $N$-tuply excited state determinants, respectively.

The coupled cluster energy and amplitude equations can be represented as

$$E_{\text{CC}} = \left\langle \Phi_0 \middle| e^{-\hat{T}} \hat{\text{H}} e^{\hat{T}} \middle| \Phi_0 \right\rangle \tag{32}$$

$$\left\langle \mu \middle| e^{-\hat{T}} \hat{\text{H}} e^{\hat{T}} \middle| \Phi_0 \right\rangle = 0 \tag{33}$$

Due to the non-Hermitian nature of coupled cluster similarity transformed Hamiltonian, the left-hand wave function is not the adjoint of $\left|\psi_{CC}\right\rangle$, instead it is parametrized as

$$\langle \tilde{\psi}_{CC} | = \langle \Phi_0 | (1 + \hat{\Lambda}) e^{-\hat{T}} \tag{34}$$

where $\hat{\Lambda}$ is a linear de-excitation operator and defined as

$$\langle \Phi_0 | \hat{\Lambda} = \langle \Phi_0 | \sum_{\mu} \hat{\Lambda}_{\mu} = \langle \Phi_0 | \sum_{\mu} \lambda_{\mu} \hat{\tau}_{\mu}^{\dagger} = \sum_{\mu} \lambda_{\mu} \langle \mu | \tag{35}$$

The $\lambda_\mu$ are the corresponding amplitudes for de-excitations and

$$\langle \mu | = \langle \phi_0 | \hat{\tau}_{\mu}^{\dagger} \tag{36}$$

Similar to the $t_\mu$ amplitude equation described in eq. (34), the $\lambda_\mu$ amplitudes can also be obtained as:

$$\left\langle \Phi_0 \middle| (1 + \hat{\Lambda}) [\bar{H}, \hat{\tau}_{\mu}] \middle| \Phi_0 \right\rangle = 0 \tag{37}$$

***(b) Time-dependent coupled cluster theory***

The time evolution of coupled cluster wave function under a time-dependent external perturbation can be denoted as

$$\left|\Psi_{CC}(t)\right\rangle = e^{\hat{T}(t)} | \Phi_0 \rangle e^{i\epsilon(t)} \tag{38}$$

and due to the non-Hermitian nature of CC Hamiltonian, it also leads to a distinct time-dependent left-hand side analogue as

$$\langle \tilde{\Psi}_{CC}(t) | = \langle \Phi_0 | (1 + \hat{\Lambda}(t)) e^{-\hat{T}(t)} e^{-i\epsilon(t)} \tag{39}$$

Where $e^{i\epsilon(t)}$ implies a time-dependent phase factor and $\hat{T}(t)$ is a time-dependent cluster operator with time-dependent amplitudes.

As the cluster operator is also time-dependent, they generate excited state determinants upon acting on the reference state with inherent time dependency in the cluster amplitudes $t_\mu(t)$ as,

$$\hat{T}(t)|\Phi_0\rangle = \sum_\mu \hat{T}_\mu(t)|\Phi_0\rangle = \sum_\mu t_\mu(t)\hat{\tau}_\mu|\Phi_0\rangle = \sum_\mu t_\mu(t)|\mu\rangle \quad (40)$$

Similarly, for the time-dependent de-excitation operator,

$$\langle\Phi_0|\hat{\Lambda}(t) = \langle\Phi_0|\sum_\mu \hat{\Lambda}_\mu(t) = \langle\Phi_0|\sum_\mu \lambda_\mu(t)\hat{\tau}_\mu^\dagger = \sum_\mu \lambda_\mu(t)\langle\mu| \quad (41)$$

It should be noted that the orbital response is not considered for the reference wave function, generated from SCF using SFX2C1e Hamiltonian. The time-dependent coupled cluster amplitudes $t_\mu(t)$ and $\lambda_\mu(t)$ are given by

$$\langle\mu|\bar{H}|\Phi_0\rangle = i\frac{dt_\mu}{dt} \quad (42)$$

$$\left\langle\Phi_0\left|(1+\hat{\Lambda})\left[\bar{H},\tau_\mu\right]\right|\Phi_0\right\rangle = -i\frac{d\lambda_\mu}{dt} \quad (43)$$

***(c) Linear response function***

The external response can be manifested by employing a time-dependent perturbation $\hat{V}^{(1)}(t)$ on the original time-independent Hamiltonian such as

$$\hat{H} = \hat{H}^{(0)} + \hat{V}^{(1)}(t) \quad (44)$$

This time-dependent operator $\hat{V}^{(1)}(t)$ can be converted into a frequency-dependent Hamiltonian $\hat{V}^{(1)}(\omega)$ using Fourier-Transformation[1] as

$$\hat{V}^{(1)}(t) = \int_{-\infty}^{\infty} d\omega \hat{V}^{(1)}(\omega)e^{-i\omega t} \quad (45)$$

and the linear response function for an exact state is given by

$$\left\langle\left\langle A,\hat{V}^{(1)}(\omega)\right\rangle\right\rangle = \sum_n \left(\frac{\left\langle\Phi_0\left|A\right|\Psi_n\right\rangle\left\langle\Psi_n\left|\hat{V}^{(1)}(\omega)\right|\Phi_0\right\rangle}{\omega-\omega_n} - \frac{\left\langle\Phi_0\left|\hat{V}^{(1)}(\omega)\right|\Psi_n\right\rangle\left\langle\Psi_n\left|A\right|\Phi_0\right\rangle}{\omega+\omega_n}\right) \quad (46)$$

Here $\Psi_n$ is the solution of the unperturbed Hamiltonian $\hat{H}^{(0)}$ and $\omega_n$ is the excitation energy corresponding to $n^{\text{th}}$ state. Now, the perturbed excitation and de-excitation cluster operators are expanded in terms of perturbation order as

$$\hat{T}(t) = \hat{T}^{(0)}(t) + \hat{T}^{(1)}(t) + \hat{T}^{(2)}(t) + \ldots \quad (47)$$

$$\hat{\Lambda}(t) = \hat{\Lambda}^{(0)}(t) + \hat{\Lambda}^{(1)}(t) + \hat{\Lambda}^{(2)}(t) + \ldots \quad (48)$$

The expression for the first-order perturbed $t_\mu^{(1)}$ amplitude can be written as:

$$\omega t_{\mu}^{(1)}(\omega) = \left\langle \mu \middle| \bar{V}^{(1)}(\omega) \middle| \phi_0 \right\rangle + \left\langle \mu \middle| \left[ \bar{H}^{(0)}, \hat{T}^{(1)}(\omega) \right] \middle| \phi_0 \right\rangle \tag{49}$$

Similarly, the perturbed amplitude equation for $\lambda_{\mu}^{(1)}$ have the following form

$$-\omega \lambda_{\mu}^{(1)}(\omega) = \left\langle \phi_0 \middle| \hat{\Lambda}^{(1)}(\omega) \left[ \bar{H}^{(0)}, \tau_{\mu} \right] \middle| \phi_0 \right\rangle + \left\langle \phi_0 \middle| \left(1 + \hat{\Lambda}^{(0)}\right) \left[ \tilde{V}^{(1)}(\omega), \tau_{\mu} \right] \middle| \phi_0 \right\rangle \tag{50}$$

where

$$\tilde{V}^{(1)}(\omega) = \bar{V}^{(1)}(\omega) + \left[ \bar{H}^{(0)}, \hat{T}^{(1)}(\omega) \right] \tag{51}$$

The time-dependent expectation value of an operator $\hat{A}$ can be written as

$$\langle \hat{A}(t) \rangle = \left\langle \phi_0 \middle| \left(1 + \hat{\Lambda}(t)\right) \bar{A} \middle| \phi_0 \right\rangle \tag{52}$$

Consequently, the coupled cluster linear response function may be represented as,

$$\left\langle \left\langle \hat{A}; \hat{V}^{(1)}(\omega) \right\rangle \right\rangle = \left\langle \phi_0 \middle| \left[ \left(1 + \hat{\Lambda}^{(0)}\right) \left[ \bar{A}^{(0)}, \hat{T}^{(1)}(\omega) \right] + \hat{\Lambda}^{(1)}(\omega) \bar{A}^{(0)} \right] \middle| \phi_0 \right\rangle \tag{53}$$

It can be seen that the linear response function, as described above, is asymmetric in nature as it involves both first-order perturbed $\hat{T}^{(1)}$ and $\hat{\Lambda}^{(1)}$ operator. One can also express the linear response function in its symmetric form by writing it only in terms of perturbed $\hat{T}^{(1)}(\omega)$ operator, which will consist of quadratic terms $\hat{T}^{(1)}(\omega)$.[55]

$$\begin{aligned} \left\langle \left\langle \hat{A}; \hat{V}^{(1)}(\omega) \right\rangle \right\rangle = &\left\langle \phi_0 \middle| \left(1 + \hat{\Lambda}^{(0)}\right) \left[ \bar{A}^{(0)}, \hat{T}_{\hat{V}^{(1)}}^{(1)}(\omega) \right] \middle| \phi_0 \right\rangle + \left\langle \phi_0 \middle| \left[ \hat{V}^{(1)}(\omega), \hat{T}_{\hat{A}}^{(1)}(-\omega) \right] \middle| \phi_0 \right\rangle \\ &+ \left\langle \phi_0 \middle| \left[ \left[ \bar{H}^{(0)}, \hat{T}_{\hat{V}^{(1)}}^{(1)}(\omega) \right], \right] \hat{T}_{\hat{A}}^{(1)}(-\omega) \middle| \phi_0 \right\rangle \end{aligned} \tag{54}$$

In this work, we have used the asymmetric formalism as described in eq (53) for the polarizability calculation. The dipole polarizability tensor within the linear response function can be represented as

$$\left\langle \left\langle \boldsymbol{\mu}_{\mathcal{A}}; \boldsymbol{\mu}_{\mathcal{B}} \right\rangle \right\rangle = \left\langle \phi_0 \middle| \left[ \left(1 + \hat{\Lambda}^{(0)}\right) \left[ \bar{\boldsymbol{\mu}}_{\mathcal{A}}, \hat{T}_{\mathcal{B}}^{(1)}(\omega) \right] + \hat{\Lambda}_{\mathcal{B}}^{(1)}(\omega) \bar{\boldsymbol{\mu}}_{\mathcal{A}} \right] \middle| \phi_0 \right\rangle \tag{55}$$

where $\bar{\boldsymbol{\mu}}$ is the similarity transformed dipole moment operator. Subsequent symmetrization and trace of the polarizability tensor finally give the mean polarizability

$$\alpha(\omega) = \frac{1}{3} Tr\left[ \left\langle \left\langle \mu; \mu \right\rangle \right\rangle \right] \tag{56}$$

***(d) Density fitting (DF) approximation***

To further reduce the storage and computational overhead, we have employed density fitting approximation where four-centered two-electron integrals are replaced by three-centered two-electron integrals which reduces the computational cost[56,57] significantly. The four-centered two-electron integrals $(pq|rs)$ for real orbitals can be defined as,[58]

$$(pq|rs) = \int dr_1 \int dr_2 \phi_p(r_1)\phi_q(r_1) \frac{1}{r_{12}} \phi_r(r_2)\phi_s(r_2) \tag{57}$$

One can rewrite the expression for two-electron integrals as,

$$(pq|rs) = \int dr_1 \int dr_2 \boldsymbol{\rho}_{pq}(r_1) \frac{1}{r_{12}} \boldsymbol{\rho}_{rs}(r_2) \tag{58}$$

Where, $\boldsymbol{\rho}_{xy} = \phi_x(r)\phi_y(r)$. An auxiliary basis can be used to fit $\boldsymbol{\rho}_{xy}$ as,

$$\bar{\boldsymbol{\rho}}_{xy}(r) = \sum_{P}^{N_{aux}} \mathbf{d}_P^{xy} \boldsymbol{\chi}_P(r) \tag{59}$$

Here, $\mathbf{d}_P^{xy}$ and $\boldsymbol{\chi}_P$ are the fitting coefficients of the auxiliary basis and auxiliary basis functions respectively. one can express $\mathbf{d}_P^{xy}$ as,

$$\mathbf{d}_P^{xy} = \sum_Q (xy|Q)[\mathbf{X}^{-1}]_{QP} \tag{60}$$

where, $(xy|Q)$ is the three-centered two-electron integral, defined as,

$$(xy|Q) = \int dr_1 \int dr_2 \phi_x(r_1)\phi_y(r_1) \frac{1}{r_{12}} \boldsymbol{\chi}_Q(r_2) \tag{61}$$

and

$$\mathbf{X}_{QP} = \int dr_1 \int dr_2 \boldsymbol{\chi}_Q(r_1) \frac{1}{r_{12}} \boldsymbol{\chi}_P(r_2) \tag{62}$$

The four-centered two-electron integral $(pq|rs)$ can be constructed from the three-centered two-electron integrals, which are fitted by an auxiliary basis[59–61],

$$
\begin{aligned}
(pq|rs) &= \int dr_1 \int dr_2 \sum_Q \mathbf{d}_Q^{pq} \chi_Q(r_1) \frac{1}{r_{12}} \chi_r(r_2) \chi_s(r_2) \\
&= \sum_Q \mathbf{d}_Q^{pq} (Q|rs) \\
&= \sum_{PQ} (pq|P)[\mathbf{X}^{-1}]_{PQ} (Q|rs) \\
&= \sum_{PQR} (pq|P)[\mathbf{X}^{-\frac{1}{2}}]_{PQ} [\mathbf{X}^{-\frac{1}{2}}]_{QR} (R|rs) \\
&= \sum_Q \left\{ \sum_P (pq|P)[\mathbf{X}^{-\frac{1}{2}}]_{PQ} \right\} \left\{ \sum_R [\mathbf{X}^{-\frac{1}{2}}]_{QR} (R|rs) \right\} \\
&= \sum_Q \mathbf{J}_{pq}^Q \mathbf{J}_{rs}^Q
\end{aligned}
\tag{63}
$$

where

$$
\mathbf{J}_{pq}^Q = \sum_P (pq|P)[\mathbf{X}^{-\frac{1}{2}}]_{PQ} \tag{64}
$$

The factorized linear response coupled cluster equations in the term three-centered two-electron intermediates are presented in the Supporting Information. The integrals up to two external indices are constructed once and stored in the disk. The integrals with three and four external indices are never constructed explicitly, and the terms involving these integrals are always constructed on the fly from three-centered two-electron integrals. It should be noted that the density fitting approximation can only reduce the formal scaling of Coulomb kind of terms unless special techniques are used[62–65]. However, the reduction in disk I/O still leads to significant speed up even for exchange terms.

A pictorial representation of the steps involved in the SFX2C1e-LRCCSD method has been shown in Figure 1.

## 3. Computational Details

All the SFX2C1e-LRCCSD polarizability calculations based on coupled cluster linear response framework are carried out using our in-house quantum chemistry software package BAGH[66]. The converged SFX2C1e Hartree-Fock coefficient and one electron and three centered two-electron integrals were taken from PySCF v2.1[67–69]. The auxiliary functions were generated automatically based on the orbital basis set in PySCF. Experimental geometry has been used for all the molecules unless specifically mentioned otherwise, and the corresponding cartesian coordinates are provided in the Supporting Information.

# 4. Results and Discussion

## *4.1. Polarizability of atoms:*

Accurate estimation of atomic polarizability is crucial for high-precision experiments. We start our discussion with the static and dynamic polarizability of three d-block atoms, i.e. Zn, Cd and Hg. Recently Gomes and co-workers[28] have performed a detailed analysis of the pole structure of the dynamic polarizability tensor for Zn, Cd, and Hg in their exact two-component linear response coupled cluster (X2C-LRCCSD) study. In this work, we have extended the analysis to non-relativistic (NR) and SFX2C1e-LRCCSD methods. In Figure 2, we have plotted the calculations done at three different levels of theories: non-relativistic LRCCSD, SFX2C1e-LRCCSD, and exact two-component LRCCSD (X2C-LRCCSD) using s-aug-Dyall.v2z basis set. The X2C-LRCCSD results were taken from the reference.[28] For Zn and Cd, the poles correspond to two $ns \rightarrow (n+1)p$ transitions (A and B denote spin-forbidden ${}^1S_0 \rightarrow {}^3P_1^0$ and spin-allowed ${}^1S_0 \rightarrow {}^1P_0$ transitions, respectively) and similarly for $ns \rightarrow (n+2)p$ transitions two poles near C and D have been observed corresponding to spin-forbidden and spin-allowed transitions, respectively. Only the first two poles corresponding to $ns \rightarrow np$ transition are present for Hg in the frequency range considered by Gomes and co-workers.[28] It can be seen that the frequency-dependent polarizability spectrum of Zn, Cd, and Hg obtained from the SFX2C1e variant closely resembles the trend calculated at the full X2C-LRCCSD level for the entire frequency range except for regions A and C. It proves that second-order response properties can be accurately simulated even with the inclusion of the scalar relativistic effects only. It is justified, as the spin-free Hamiltonian for closed-shell atoms and molecules contributes at the first order in perturbation to the relativistic Hamiltonian, while the spin-dependent part of the Hamiltonian appears in the higher orders of perturbation. The non-relativistic LRCCSD results show considerable deviation from the SFX2C1e-LRCCSD and X2C-LRCCSD values, and the deviation between relativistic and non-relativistic results is more prominent at higher frequencies. Although the SFX2C1e-LRCCSD and X2C-LRCCSD give excellent agreement for most of the frequency ranges and both of them have almost similar qualitative picture of polarizability spectrum, but the former misses the poles corresponding to spin-forbidden excited states present in the X2C-LRCCSD. For instance, when considering the element Zn, the peaks obtained at frequencies 0.140 a.u. (A) and 0.27 a.u. (C) in X2C-LRCCSD calculations are not observed in SFX2C1e and non-relativistic results. These poles correspond to the spin-forbidden ${}^1S_0 \rightarrow {}^3P_1$ transitions arising due to the spin-orbit coupling (SOC), which cannot be captured by the SFX2C1e method where the spin-dependent terms are absent. The SFX2C1e-LRCCSD method, however, accurately reproduces the poles at ~0.2089 a.u. (B) and ~0.2828 a.u. (D) which arise due to spin-allowed ${}^1S_0 \rightarrow {}^1P_0$ transitions. A similar pattern is observed for Cd and Hg atoms, where the peaks due to spin-forbidden transitions cannot be captured by the SFX2C1e-LRCCSD model.

***(a) Effect of density fitting (DF):***

To understand the magnitude of the error caused by density fitting approximation, we have plotted the dynamic polarizability value of Zn, Cd, and Hg for the frequency range 0.0-0.3 a.u. with and without density fitting in Figure 3. The s-aug-Dyall.v2z basis set has been used for the calculations. It can be seen that the use of density fitting approximation leads to very negligible error in the SFX2C1e-LRCCSD calculations, and the plots with and without density fitting are almost indistinguishable from one another. Therefore, the rest of the calculations in the manuscript are performed employing density fitting approximation.

***(b) Effect of relativity and electron correlation:***

Table 1 presents the effect of relativity and correlation on static and dynamic polarizability. s-aug-Dyall.v2z basis set has been used for the calculations. The Hartree-Fock polarizabilities were calculated using the sum over state (SOS)[70] formula using TDHF[71–74] energy and wave function. It can be seen that the effect of correlation and relativity both lead to the reduction of the polarizability values.  The relativistic effect has a small impact on the static polarizability of Zn, and the error in NR-LRCCSD result is 6 percent with respect to its SFX2C1e version. As we move down the group, the error steadily grows, and neglecting the relativistic effect can result in significant errors of up to 19% and 69% for Cd and Hg, respectively. Even in the case of Zn, the impact of relativity on dynamic polarizability is considerable, with a notable 12 percent disparity observed between relativistic and non-relativistic LRCCSD at a frequency of 0.14014 a.u.. As the frequency increases, the electron correlation starts to show a greater impact on the dynamic polarizability. The dynamic polarizability (at 0.14014 a.u. frequency) of Zn is reduced by 38 percent when electron correlation is taken into account, compared to the SFX2C1e Hartree-Fock method. It can be seen that effect correlation is more prominent in the non-relativistic case than in the relativistic CCSD method. The effect of correlation is most prominent for dynamic polarizability of Hg at the frequency 0.14014a.u., where one can observe a reduction of 77 percent on going from the non-relativistic Hartree-Fock to the non-relativistic LRCCSD method. For the same frequency, the corresponding reduction is 35 percent in the case of relativistic calculation, which is almost half of the non-relativistic one.

***(c) Effect of basis set:***

The basis set plays a vital role in determining the accuracy of the calculated polarizability values. To understand the effect of the basis set on the SFX2C1e-LRCCSD polarizability values, we have investigated the basis set dependence on the dynamic polarizability values on Kr atom. The experimental value[75] is available for the dynamic polarizability of Kr at 0.072 a.u. frequency. Therefore, we have selected that particular frequency for our calculations. We have used Dyall.aenz (n=2,3,4) relativsitic basis sets.[76–84] It is well known that accurate polarizability calculations require the inclusion of sufficient diffuse functions in the basis set[85]. To see the effect of diffuse functions on the calculation, we have augmented the basis set with single, double, triple, and quadruple sets of diffuse functions. These augmented basis sets are generated using DIRAC19[86,87] software package.

From Table 2, it can be observed that the absence of diffuse functions leads to considerable error in SFX2C1e-LRCCSD results. For example, the deviation in SFX2C1e-LRCCSD results in the Dyall.ae2z basis set amounts to 9.737 a.u. when compared to the experimental values. The error reduces to 4.966 a.u. when the Dyall.ae3z basis set is used. The error does not converge even when the Dyall.ae4z basis set is used. But even a single augmentation at Dyall.ae2z level reduces the error to 0.404 a.u. The results generally converge with the triple augmentation of the basis set at double and triple zeta levels, whereas single augmentation is sufficient at quadruple zeta levels. The triply augmented triple zeta results show better agreement with the experiment than the corresponding quadrupole zeta basis set, presumably due to fortuitous error cancelation. Tables S1 and S2 present the polarizability values of group 2 and group 18 elements in t-aug-Dyall.ae3z and s-aug-Dyall.ae4z basis set. It can be seen that the results in these two basis sets are generally in close agreement with each other, similar to that observed for the Kr atom. The results are also in good agreement with the available experimental results. Therefore, the s-aug-Dyall.ae4z can be taken as an optimal basis set for the calculation.

As we have seen, the basis set plays a crucial role in determining the accuracy of the calculated polarizabilities, especially when one is interested in comparing with experimental results. Table 3 presents the experimental and theoretical values of static and dynamic polarizabilities for Zn, Cd, and Hg. The calculations were performed using the s-aug-Dyall.ae4z basis set at the SFX2C1e-LRCCSD level of theory. We have also included the X2C-LRCCSD results by Gomes and co-workers[28] in our comparison. The static and dynamic polarizabilities determined using the SFX2C1e-LRCCSD method show good agreement with the experimental values. The SFX2C1e-LRCCSD static polarizability values overestimate the experimental results by 0.54 a.u. and 1.12 a.u for Zn and Hg respectively. On the other hand, the static polarizability value for Cd is underestimated by 0.69 a.u. This slight deviation may be attributed to the contribution from spin-orbital coupling and triples or higher-order correlation corrections missing in the present implementation. The non-relativistic static polarizability value for Hg shows an error of 72.6% with respect to the experiment. This demonstrates the importance of the inclusion of the relativistic effect for polarizability calculations of heavy elements. The deviation between the experimental values and the SFX2C1e-LRCCSD static polarizability values is smaller than those reported by Gomes and co-workers[28] for the X2C-LRCCSD method. This could be due to the use of a smaller s-aug-Dyall.v2z basis set in their calculations[28]. Table 3 also presents the dynamic polarizability values at three different frequencies. The SFX2C1e-LRCCSD method for Zn shows 1.65%, 1.83%, and 1.90% of error with respect to experiment for dynamic polarizabilities at frequencies 0.07198 a.u., 0.08383 a.u., and 0.14014 a.u., respectively. Slightly higher errors are observed for Cd and Hg, which is presumably due to the higher magnitude of spin-orbital coupling. Conversely, the non-relativistic LRCCSD method fails to accurately reproduce the dynamic polarizabilities, even in a qualitative sense. For Zn, the non-relativistic dynamic polarizability value at a frequency of 0.14014 a.u. exhibits an error of 15.36% relative to the experiment. The error can reach as high as 188.1% for Hg at the same frequency.

### *4.2. Polarizability of molecules:*

Molecular polarizability can be different from atomic polarizability as the molecule lacks the spherical symmetry of atoms. One can calculate the mean polarizability of the molecule as

$$\bar{\alpha}(\omega) = \frac{\alpha_{xx}(\omega) + \alpha_{yy}(\omega) + \alpha_{zz}(\omega)}{3} \tag{65}$$

and the anisotropic polarizability can be described as

$$\gamma(\omega) = \alpha_{zz}(\omega) - \frac{1}{2}\left(\alpha_{xx}(\omega) + \alpha_{yy}(\omega)\right) \tag{66}$$

where, $\alpha_{zz}(\omega)$ is the parallel component and $\alpha_{xx}(\omega)$, $\alpha_{yy}(\omega)$ are known as perpendicular components for a particular frequency $\omega$, when the external electric field is directed along the z-axis.

In Table 4, we have presented the static polarizability of AuH. Since no experimental data for the static polarizability is available for AuH, we have compared the SFX2C1e-LRCCSD values with the previously reported 4c-DFT linear response results by *Salek et al*[21]. The double and triple zeta quality basis set chosen in the present work is slightly different from the 4c-DFT calculations. In the 4c-DFT calculation, the used basis sets were of dimensions 22*s*19*p*12*d*9*f* and 29*s*24*p*15*d*11*f*. On the other hand, we employed Dyall.ae2z and Dyall.ae3z basis sets with dimensions *24s19p12d9f1g* and *30s24p15d11f5g1h*, respectively. For hydrogen, *Salek et al* have used Dunning's de-contracted aug-cc-pVXZ (X=D, T) basis set, but we used the same in its contracted form. We have also calculated SFX2C1e-LRCCSD static polarizability values on a quadruple zeta quality (QZ) basis, where s-aug-Dyall.ae4z is used for Au and uncontracted aug-cc-pVQZ for H. The mean ($\bar{\alpha}$), parallel ($\alpha_{//}$), and perpendicular ($\alpha_{\perp}$) polarizability components calculated at the SFX2C1e-LRCCSD level agree qualitatively with the DFT values for all three functionals. The absolute numerical values differ from functional to functional, but the increase in the polarizability value in SFX2C1e-LRCCSD method with the increase of basis set size is also a common trend observable for all the functionals.

In Table 5, we have presented static polarizabilities of hydrogen halides (HX, where X=F, Cl, Br, I), iodine dimer, and molecules made of alkali metals, i.e. NaLi and CsK. Gomes and co-workers[28] used the same set of molecules to benchmark their X2C-LRCCSD implementation. To demonstrate the performance of the SFX2C1e-LRCCSD method, we have compared our results with the X2C-LRCCSD variant. To separate out the relativistic effect from the basis set effect, we have also reported polarizability values at the same basis set that Gomes and co-workers[28] have used for the isotropic static polarizability calculation of diatomic molecules, i.e. uncontracted aug-cc-pVDZ for lighter elements (H, F, Cl, Br,Na,Li) and s-aug-Dyall.v2z basis set for heavier elements (I, Cs). The SFX2C1e-LRCCSD method shows a difference of 0.08 a.u., 0.02 a.u., 0.08 a.u for HF, HCl, and HBr, respectively, with respect to the X2C-LRCCSD method. A slightly higher deviation from X2C results was observed for HI, ICl, and $I_2$, ranging from 0.3 a.u. to 0.5 a.u. This is presumably due to the higher effect of spin-orbit coupling present in the molecules containing iodine. In the case of metal dimers, NaK and CsK show a much higher magnitude of deviation, where SFX2C1e-LRCCSD underestimates the static polarizability by 4 a.u. with respect to the X2C-LRCCSD result. It is not appropriate to compare the results in a double zeta quality basis set with the experimental results. The use of a larger basis set may be extremely expensive in X2C or four-component Dirac-Coulomb Hamiltonian. However, the density fitting-based implementation of SFX2C1e-LRCCSD in a spin-adapted formulation allows one to use larger basis sets in the calculations.

Figure 4 presents the error in static polarizability values with respect to the experiment in SFX2C1e and X2C-LRCCSD. The uncontracted aug-cc-pVDZ basis set has been used for the lighter elements (H, F, Cl, Br) and s-aug-Dyall.v2z basis set has been used for heavier elements (I, Cs). It can be seen that the SFX2C1e-LRCCSD method shows a very similar error with respect to the experiment as that of the X2C-LRCCSD method in the double zeta basis set. In addition to s-aug-Dyall.ae4z basis set, we have also performed these calculations in another quadruple zeta quality basis set with a comparable number of basis functions, i.e. d-aug-Dyall.v4z, to confirm the validation of the similar range of accuracy for polarizability values in these basis sets. The agreement of SFX2C1e-LRCCSD results with the experiment improves when going to the d-aug-Dyall.v4z or s-aug-Dyall.ae4z basis set for all molecules, except ICl and NaLi. It shows that the basis set has a more significant effect than the spin-orbit coupling for most of the molecules, at least for the static polarizabilities. The static polarizability values for the ICl and NaLi in the SFX2C1e-LRCCSD method in a double zeta quality basis set are very similar to that of the X2C method. However, the discrepancies between the experimental values and the calculated results become larger when using either d-aug-Dyall.v4z or s-aug-Dyall.ae4z basis set for ICl and NaLi. The reason for the increased discrepancy with respect to the experiment with the increase in the basis set dimension for these two molecules is not very clear at this moment, particularly for these two aforementioned molecules.

Experimental dynamic polarizability of iodine dimer is available from the work of Maroulis *et al.*[88] They measured the polarizability of $I_2$ at three different frequencies: 0.07198 a.u., 0.07669

a.u., and 0.140114 a.u. Gomes and co-workers[28] have reported X2C-LRCCSD dynamic polarizabilities at these frequencies using s-aug-Dyall.v2z basis set. We have also reported the dynamic polarizability results at these three frequencies using the SFX2C1e-LRCCSD level of theory. From Table 6, it can be seen that the experimental polarizability of $I_2$ increases with the frequency, and it is gratifying to note that a similar trend is also observed for the SFX2C1e-LRCCSD results. The X2C-LRCCSD dynamic polarizabilities increase from frequency 0.07198 a.u. to 0.07669 a.u. but decreases from 0.07669 a.u. to 0.14014 a.u. Table 6 also presents the relativistic polarizabilities at HF and DFT levels based on the B3LYP functional. For all three levels of theory, the parallel component $\alpha_{xx}(\omega)$ does not vary significantly with the frequency and it is also observed that the effect of correlation is not that prominent for $\alpha_{xx}(\omega)$, whereas $\alpha_{zz}(\omega)$ is significantly sensitive to the electronic correlation effect and the level of theory. One can see that the parallel component shows a negative value at 0.07198 a.u. and 0.07669 a.u. for HF and DFT, respectively, signifying that near these two frequencies, they have poles in the polarizability spectrum, whereas no negative polarizability is observed either for SFX2C1e-LRCCSD or X2C-LRCCSD level of calculation. From Table 5, it is observed that the mean static polarizability values computed for diatomic molecules using s-aug-Dyall.ae4z and d-aug-Dyall.v4z basis set provide almost comparable accuracy. So, to compare with experimental results we have also provided the dynamic polarizabilities of $I_2$ for the three aforementioned frequencies using d-aug-Dyall.v4z basis set. Both $\alpha_{xx}(\omega)$ and $\alpha_{zz}(\omega)$ in SFX2C1e-LRCCSD method increases on going to a larger d-aug-Dyall.v4z basis set. It also leads to an improved agreement of the mean polarizability value with the available experimental result.

The density fitting-based implementations allow one to apply the SFX2C1e-LRCCSD method beyond diatomic molecules**.** We have selected three molecules, namely Uranium Hexafluoride ($UF_6$), Osmium Tetraoxide ($OsO_4$), and Mercury Chloride ($HgCl_2$), for which experimental static polarizability values are available. Previous theoretical estimations of static polarizability for these three molecules are reported by Cremer and co-workers[51] using analytical second derivatives of energy based on the Normalized Elimination of Small Component (NESC) framework. In Table 7, we have included the computed static polarizability values obtained using the SFX2C1e-LRCCSD approach, as well as the DFT and MP2 values as reported by Cremer and co-workers[51]. Additionally, the table includes experimental values for comparison. We have used SARC-DKH (U, Os, and Hg) and def2-QZVPP (F, O, and Cl) basis set, which was also used by Cremer and co-workers[51]. The perpendicular and parallel components of $UF_6$ and $OsO_4$ calculated at SFX2C1e-LRCCSD level are exactly the same as expected from symmetric octahedral and tetrahedral symmetry of the molecular structures, respectively. On the other hand, $HgCl_2$ has different values of parallel and parapedicular components due to its linear structure. An intriguing observation from Table 7 pertains to the discrepancy between the theoretical and experimental values. Among the three molecules, $OsO_4$ shows a good agreement with the experimental value, but $UF_6$ and $HgCl_2$ results largely deviate from the available experimental values with an underestimation of 30.63 a.u. and 23.4 a.u. respectively in SFX2C1e-LRCCSD method. A similar disagreement with the experiment is also observed for the previously reported[51] DFT and MP2 results. It should be noted that the experimental numbers provided here contain both electronic and vibrational contributions to the total mean polarizability of these molecules. In the case of OsO4, the electronic contribution of the total experimental mean polarizability[89] is 51.0 ± 1.0 a.u., and the SFX2C1-1e-LRCCSD results are in excellent agreement with an overestimation by 0.45 a.u.. Meanwhile, MP2 and DFT results for the electronic component deviate by 3.54 and 5.5 a.u., respectively. The estimated electronic contribution[89] to the total polarizability for $UF_6$ is 66.6±2.0 a.u. It can be seen that the SFX2C1e-LRCCSD method shows a large deviation of 14.21 a.u for UF6. Similarly, when considering DFT calculations, a deviation of 15.25 a.u. is observed. The deviation is 12.41 a.u. at the MP2 level. This deviation from the experiment in the case of UF6 for all the levels of theories indicates a need for reconsideration and further refinement of the experiment for the static polarizability values of $UF_6$. The experimental mean polarizability of $HgCl_2$ is 78.2 a.u. However, the estimates for the individual electronic and vibrational contributions to the polarizability of $HgCl_2$ vary significantly[89] with the electronic contribution ranging from 57 a.u. to 75 a.u., and the vibrational contribution ranging from 3 a.u. to 21 a.u. The SFX2C1e-LRCCSD results show a decent agreement with the lower range of the electronic contribution.

## 5. Conclusions

We present the theory, implementation, and benchmarking of a spin-free exact two-component linear response coupled cluster method for static and dynamic polarizabilities. The SFX2C1e-LRCCSD method can make a balanced inclusion of relativistic and electron correlation effects. Density fitting approximation has been used to reduce the computational cost of the calculations, allowing one to calculate the static and dynamic polarizability at the SFX2C1e-LRCCSD level of theory beyond diatomic molecules. The SFX2C1e-LRCCSD method gives excellent agreement with the X2C-LRCCSD result at a much smaller computational cost. Our calculation shows that using a large basis set is as important as including electron correlation and relativistic effects for accurate simulation of polarizabilities of atoms and molecules containing heavy elements. The spin-free X2C treatment is sufficient in most cases for static polarizability calculations. The dynamic polarizability values are more sensitive to the electron correlation and relativistic effects, and their importance increases with the increase in frequency. The SFX2C1e-LRCCSD method accurately reproduces the pole structure in the dynamic polarizability spectrum corresponding to spin-allowed transitions but misses the poles corresponding to the spin-forbidden transitions. A potential solution to address the computational cost and/or absence of spin-forbidden poles in the polarizability spectrum, which affects the currently available relativistic LRCCSD implementations, is to develop a lower scaling LRCCSD method using a four-component relativistic Hamiltonian. Work is in progress towards that direction.

## Supporting Information

The programmable expressions for LRCCSD in the spin-adapted formulation, the static and dynamic polarizabilities of alkaline-earth metals (Be, Mg, Ca, Sr, Ba, Ra) and inert elements (He, Ne, Ar, Kr, Xe, Rn), and the used cartesian coordinates (in Å) of HF, HCl, HBr, HI, $I_2$, ICl, NaLi, CsK, AuH, $HgCl_2$, $OsO_4$, and $UF_6$ are presented in the Supporting Information.

## ACKNOWLEDGMENTS

The authors acknowledge the support from IIT Bombay, IIT Bombay Seed Grant (Project No. R.D./0517-IRCCSH0-040), CRG (Project No. CRG/2022/005672) and MATRICS (Project No. MTR/2021/000420) project of DST-SERB, CSIR-India (Project No. 01(3035)/21/EMR-II), DST-Inspire Faculty Fellowship (Project No. DST/INSPIRE/04/2017/001730), Prime Minister's Research Fellowship (PMRF) and ISRO for financial support, IIT Bombay super computational facility, and C-DAC Supercomputing resources (Param Smriti and Param Bramha) for computational time.

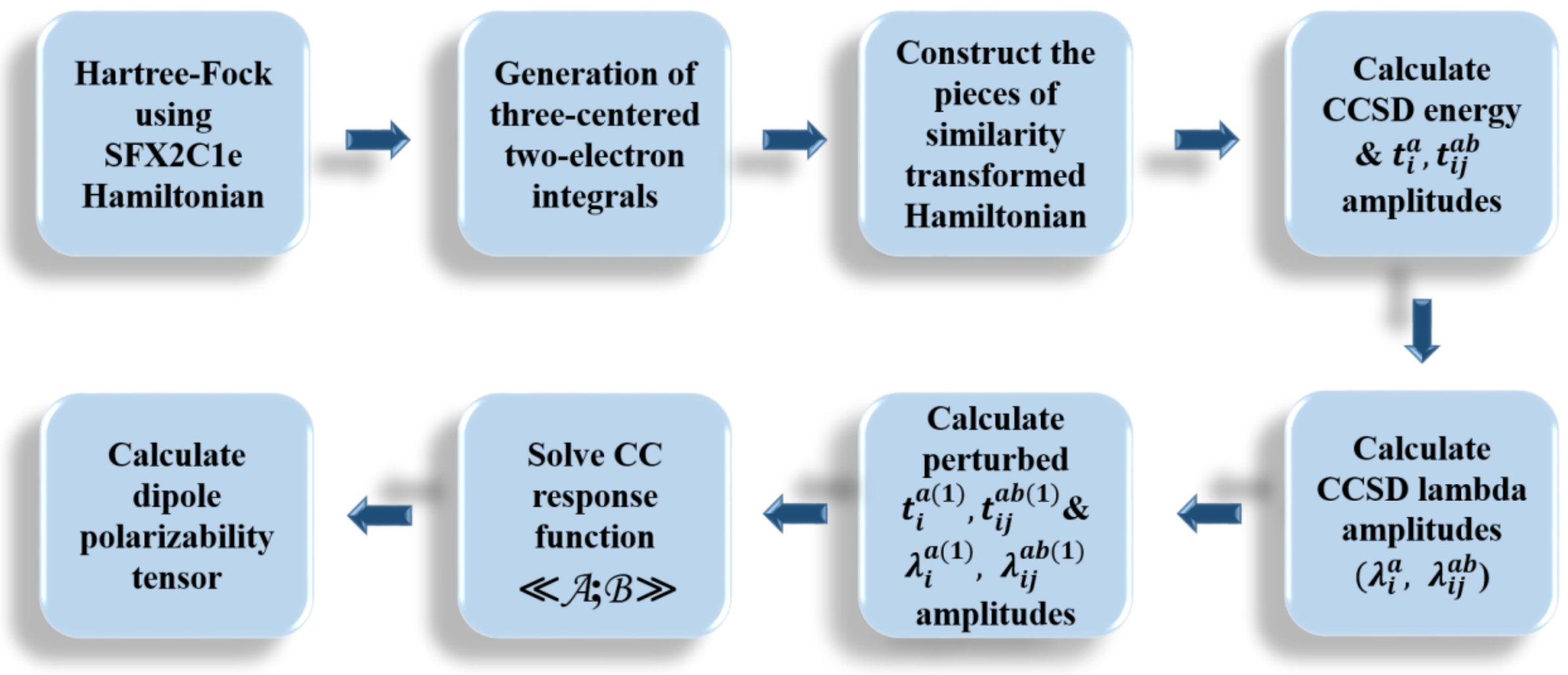

***Figure 1: The flowchart for SFX2C1e linear response coupled cluster theory***

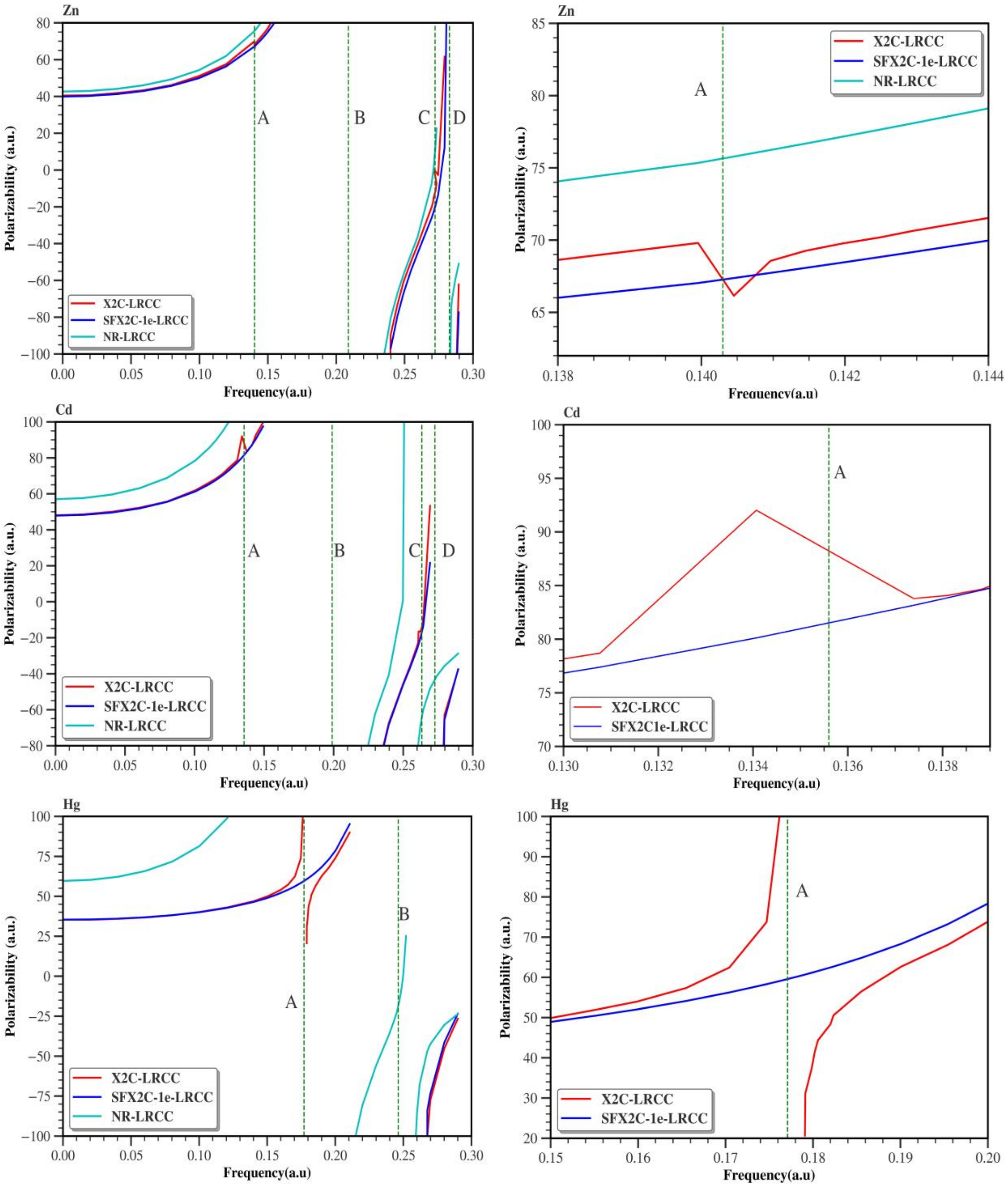

***Figure 2: The pole structure of frequency-dependent dynamic polarizabilities (in a.u.) (left) and zoomed-in version of the pole corresponding to $^1S_0 \rightarrow {}^3P_1$ transition. The X2C-LRCC results were taken from reference[28]. The s-aug-Dyall.v2z basis set has been used for the calculation***

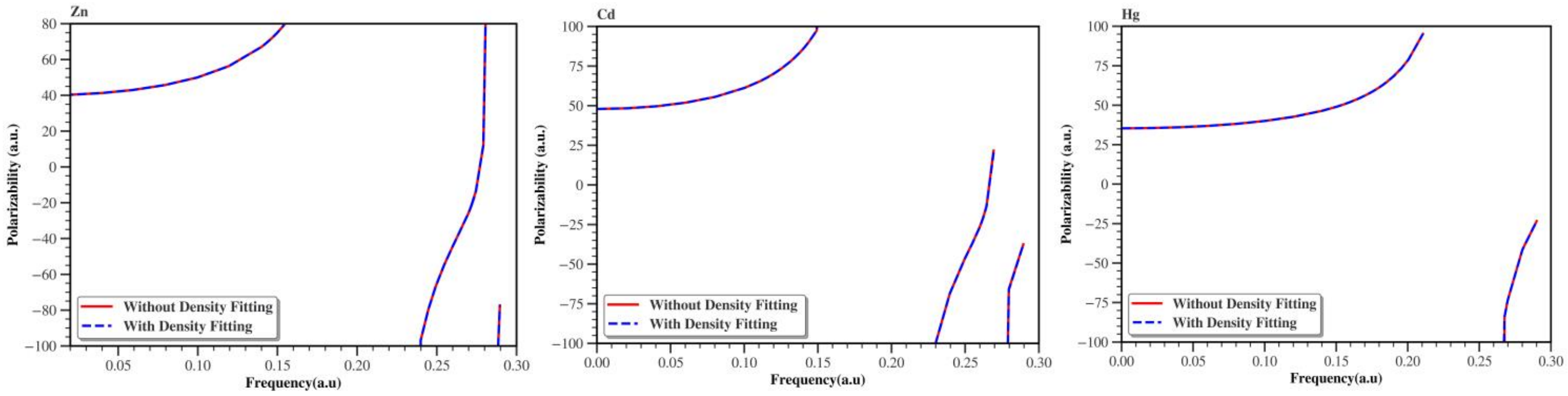

***Figure 3: SFX2C1e-LRCCSD dynamic polarizabilities (in a.u.) with and without density fitting for Zn, Cd and Hg. The s-aug-Dyall.v2z basis set has been used for the calculations.***

***Table 1: SCF and correlation contributions to the static and dynamic polarizabilities (in a.u.) in s-aug-Dyall.v2z basis set***

| | | **Polarizability** | | | |
|---|---|---|---|---|---|
| **Atom** | **Freq (a.u)** | **0.0000** | **0.07198** | **0.08383** | **0.14014** |
| Zn | NR-TDHF | 53.11 | 62.29 | 66.45 | 126.28 |
| | SFX2C1e-TDHF | 49.89 | 57.87 | 61.42 | 108.88 |
| | NR-CC | 42.64 | 47.94 | 50.20 | 75.50 |
| | SFX2C1e-CC | 39.99 | 44.57 | 46.50 | 67.15 |
| | Expt. | 38.80(8)[90] | 43.03(32)[90] | 44.76(31)[90] | 63.26(12)[90] |
| Cd | NR-TDHF | 74.91 | 93.84 | 103.40 | 404.86 |
| | SFX2C1e-TDHF | 62.79 | 75.10 | 80.91 | 186.14 |
| | NR-LRCC | 57.03 | 66.13 | 70.22 | 127.54 |
| | SFX2C1e-LRCC | 47.89 | 53.81 | 56.36 | 85.94 |
| | Expt. | 47.50 ± 2[91] | 54.20 ± 0.95[92] | 56.23 ± 0.38[92] | 68.80 ± 2.3[92] |
| Hg | NR-TDHF | 79.71 | 100.84 | 111.80 | 590.35 |
| | SFX2C1e-TDHF | 44.21 | 48.76 | 50.68 | 71.71 |
| | NR-LRCC | 59.59 | 68.87 | 73.06 | 132.84 |
| | SFX2C1e-LRCC | 35.3278 | 37.55 | 38.43 | 46.42 |
| | Expt. | 33.92(7)[93] | 35.75(310)[93] | 36.63(317)[93] | 44.64(331)[93] |

***Table 2: Dynamic polarizability (in a.u.) and benchmarking of basis set for SFX2C1e-LRCCSD calculations for Kr atom at frequency 0.072 a.u.***

| Augmentation | Dyall.ae2z | Dyall.ae3z | Dyall.ae4z | Expt. |
|---|---|---|---|---|
| without aug | 7.338 | 12.109 | 14.548 | |
| s-aug | 16.671 | 17.108 | 16.984 | |
| d-aug | 17.314 | 17.052 | 16.994 | 17.075[75] |
| t-aug | 17.321 | 17.115 | 16.995 | |
| q-aug | 17.326 | 17.115 | 16.995 | |

***Table 3: Comparison of non-relativistic (NR), SFX2C1e and X2C-LRCCSD results for static and dynamic polarizabilities (in a.u.) with experimental results***

| Atom | Frequency | Polarizability | | | |
|---|---|---|---|---|---|
| | | SFX2C1e[a] | NR[a] | X2C[b] | Expt. |
| Zn | 0.00000 | 39.34 | 41.81 | 40.42 | 38.80[90] |
| | 0.07198 | 43.74 | 46.89 | | 43.03[90] |
| | 0.08383 | 45.58 | 49.05 | | 44.76[90] |
| | 0.14014 | 65.16 | 72.98 | | 63.26[90] |
| Cd | 0.00000 | 46.81 | 55.73 | 48.25 | 47.50[91] |
| | 0.07198 | 52.40 | 64.41 | | 54.20[92] |
| | 0.08383 | 54.79 | 68.31 | | 56.23[92] |
| | 0.14014 | 82.05 | 120.59 | | 68.80[92] |
| Hg | 0.00000 | 35.04 | 58.55 | 35.25 | 33.92[93] |
| | 0.07198 | 37.22 | 67.55 | | 35.75[93] |
| | 0.08383 | 37.77 | 71.61 | | 36.63[93] |
| | 0.14014 | 45.37 | 128.61 | | 44.64[93] |

[a] s-aug-Dyall.ae4z
[b] s-aug-Dyall.v2z, The X2C-LRCC results were taken from reference[28]

***Table 4: Static polarizabilities (in a.u.) of AuH in SFX2C1e-LRCCSD and 4C-LR-DFT[21] method***

| Method | Basis Set Quality | $\bar{\alpha}$ | $\alpha_{//}$ | $\alpha_{\perp}$ |
|---|---|---|---|---|
| SFX2C1e-LRCCSD | QZ[a] | 37.82 | 40.83 | 36.32 |
| SFX2C1e-LRCCSD | DZ[b] | 36.33 | 40.29 | 34.36 |
| | TZ[b] | 37.31 | 40.47 | 35.74 |
| 4c-DFT (B3LYP)[c] | uncDZ | 35.67 | 38.63 | 34.19 |
| | uncTZ | 36.76 | 38.99 | 35.64 |
| 4c-DFT (BLYP)[)c] | uncDZ | 35.99 | 38.97 | 34.50 |
| | uncTZ | 37.15 | 39.32 | 36.07 |
| 4c-DFT (LDA)[c] | uncDZ | 34.97 | 37.91 | 33.49 |
| | uncTZ | 36.36 | 38.53 | 35.28 |

[a]Au: s-aug-Dyall.ae4z, H: unc-aug-cc-pVQZ
[b]Au: Dyall.aenz, H: aug-cc-pVXZ
[c] Taken from *Salek et al*[21]

***Table 5: Static polarizabilities (in a.u.) of diatomic molecules in different variants of two-component relativistic LRCCSD method***

| Molecule | X2C[a] | X2C[b] | SFX2C1e[c] | SFX2C1e[d] | SFX2C1e[e] | Expt |
|---|---|---|---|---|---|---|
| HF | 5.05 | 5.52 | 5.58 | 4.97 | 5.53 | 5.60±0.10[94] |
| HCl | 16.09 | 17.14 | 17.34 | 16.11 | 17.18 | 17.39±0.20[94] |
| HBr | 22.58 | 24.02 | 23.82 | 22.20 | 23.60 | 23.74±0.50[94] |
| HI | 34.30 | | 35.42 | 34.76 | 35.12 | 35.30±0.50[95] |
| ICl | 47.59 | | 49.18 | 48.08 | 48.91 | 43.80±4.40[96] |
| $I_2$ | 69.72 | | 70.53 | 69.41 | 70.11 | 69.70±1.80[88] |
| NaLi | 240 | 239 | 229 | 236 | 229 | 263±20[97] |
| CsK | 611 | | 601 | 607 | 599 | 601±44[98] |

a. Relativistic calculation using the full X2C Hamiltonian (s-aug-Dyall.v2z)[28]
b. Using diffuse triple-zeta basis set based on full X2C Hamiltonian[28]
c. SFX2C1e using d-aug-Dyall.v4z basis set (This work)
d. SFX2C1e using s-aug-Dyall.v2z basis set (I, Cs) and unc-aug-cc-pVDZ (H, F, Cl, Br) (This work)
e. SFX2C1e using s-aug-Dyall.ae4z basis set (This work)

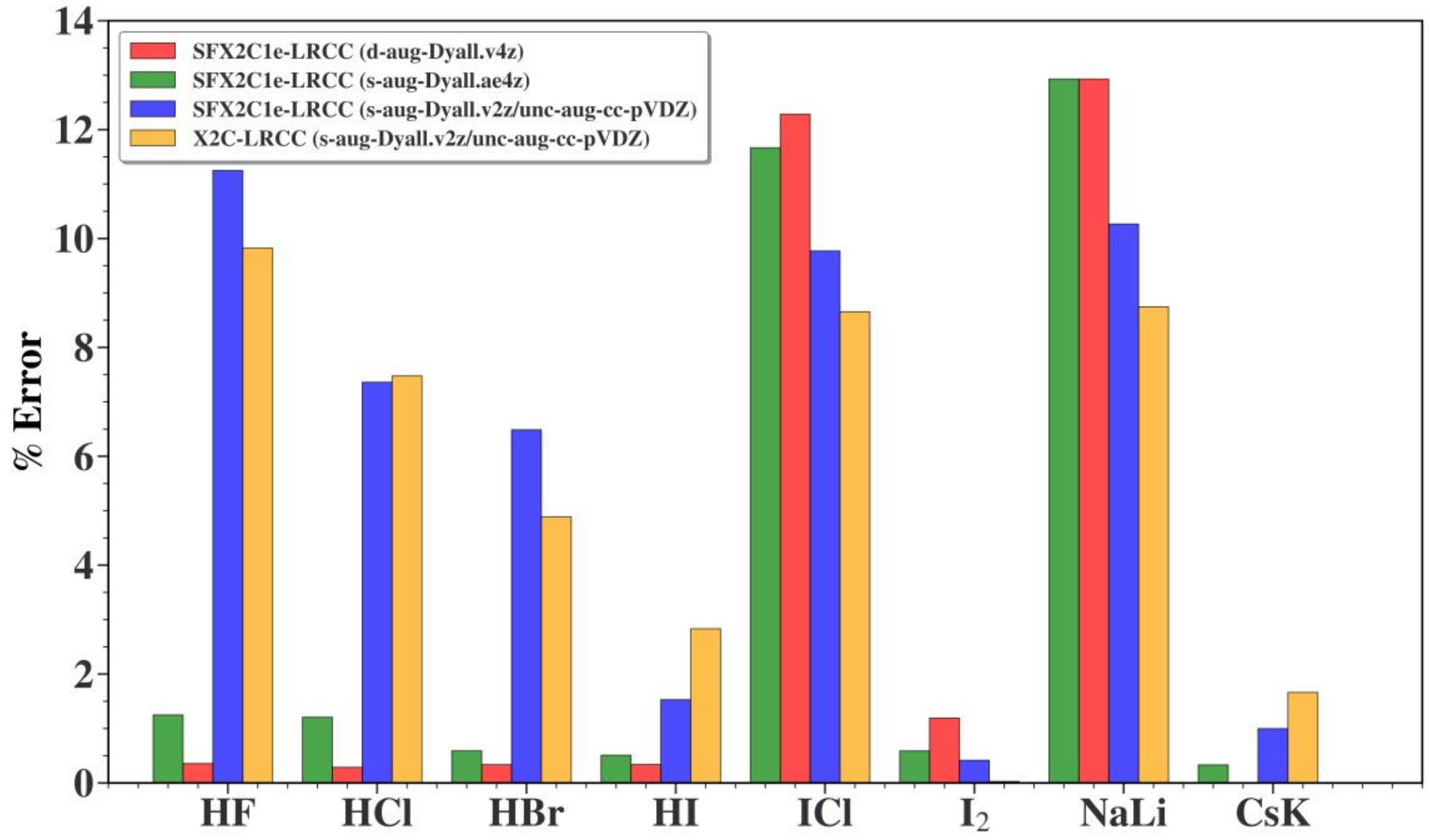

***Figure 4: Percentage error in LRCCSD static polarizability values with respect to the experimental results.***

***Table 6: Dynamic polarizabilities (in a.u.) of Iodine dimer ($I_2$)***

| | $\alpha_{xx}(\omega)$ | $\alpha_{zz}(\omega)$ | $\alpha(\omega)$ |
|---|---|---|---|
| Frequency = 0.07198 a.u. | | | |
| X2C-HF[a28] | 55.0 | 152.0 | 87.4 |
| B3LYP[a28] | 58.7 | -10.7 | 35.6 |
| X2C- LRCCSD [a28] | 55.8 | 114.8 | 75.5 |
| SFX2C1e-LRCCSD[b] | 58.55 | 104.14 | 73.7 |
| SFX2C1e-LRCCSD[c] | 57.99 | 101.97 | 72.6 |
| Expt.[88] | | | 86.8 $\pm$ 2.2 |
| Frequency = 0.07669 a.u. | | | |
| X2C-HF[a] | 56.0 | -97.3 | 4.9 |
| B3LYP[a] | 62.0 | 75.4 | 66.5 |
| X2C- LRCCSD [a] | 56.8 | 124.0 | 79.2 |
| SFX2C1e- LRCCSD[b] | 59.1 | 104.8 | 74.3 |
| SFX2C1e- LRCCSD[c] | 58.70 | 102.6 | 73.34 |
| Expt.[88] | | | 93.6 $\pm$ 3.4 |
| Frequency = 0.14014 a.u | | | |
| X2C-HF[a] | 55.3 | 117.9 | 76.2 |
| B3LYP[a] | 61.0 | 114.5 | 78.8 |
| X2C- LRCCSD [a] | 59.9 | 113.9 | 77.9 |
| SFX2C1e- LRCCSD[b] | 62.4 | 121.2 | 82.0 |
| SFX2C1e- LRCCSD[c] | 60.5 | 118.6 | 79.8 |
| Expt.[88] | | | 95.3 $\pm$ 1.9 |

[a] HF, X2C, and B3LYP results are taken from Yuan *et al.*[28]
[b] d-aug-Dyall.v4z
[c] s-aug-Dyall.v2z

*Table 7: Static polarizabilities (a.u.) of $UF_6$, $OsO_4$ and $HgCl_2$*

| molecule | Method | $\alpha_{xx}(\omega)$ | $\alpha_{yy}(\omega)$ | $\alpha_{zz}(\omega)$ | $\alpha(\omega)$ |
|---|---|---|---|---|---|
| $UF_6$ | SFX2C1e-LRCCSD[a] | 52.39 | 52.39 | 52.39 | 52.39 |
| | NESC/PBE0[51] | 51.35 | 51.35 | 51.35 | 51.35 |
| | NESC/MP2[51] | 54.19 | 54.19 | 54.19 | 54.19 |
| | Expt.[89] | | | | 83.02 ± 0.5 |
| $OsO_4$ | SFX2C1e-LRCCSD[a] | 52.45 | 52.45 | 52.45 | 52.45 |
| | NESC/PBE0[51] | 46.50 | 46.50 | 46.50 | 46.50 |
| | NESC/MP2[51] | 55.54 | 55.54 | 55.54 | 55.54 |
| | Expt.[89] | | | | 55.1 ± 0.8 |
| $HgCl_2$ | SFX2C1e-LRCCSD[a] | 42.10 | 42.10 | 80.21 | 54.80 |
| | NESC/PBE0[51] | 40.76 | 40.76 | 91.91 | 57.83 |
| | NESC/MP2[51] | 40.42 | 40.42 | 93.26 | 58.03 |
| | Expt.[89] | | | | 78.2 ± 1 |

[a]SARC-DKH: U, Os, Hg ;   def2-QZVPP: F, O, Cl

# Supporting Information

## Spin-free exact two-component linear response coupled cluster theory for estimation of frequency-dependent second-order property

Sudipta Chakraborty[a], Tamoghna Mukhopadhyay[a] and Achintya Kumar Dutta*,[a]

[a]*Department of Chemistry, Indian Institute of Technology Bombay, Mumbai-400076, India*

**achintya@chem.iitb.ac.in*

# Contents

***Table S1: Static and dynamic polarizabilities (in a.u.) of Group-2 atoms***

| Atom | Freq. (a.u.) | SFX2C1e-LRCC[a] | SFX2C1e-LRCC[b] | Ref. |
|---|---|---|---|---|
| Be | 0.070 | 43.29 | 43.68 | 43.26[1] |
| Mg | 0.000 | 71.38 | 71.42 | 71.5±3.5[2] |
| Ca | 0.000 | 156.37 | 156.12 | 169±17[3] |
| Sr | 0.000 | 196.06 | 195.76 | 186±15[4] |
| Ba | 0.000 | 269.47 | 270.09 | 268±22[3] |
| Ra | 0.000 | | 245.75 | 236±15[5]* |

[a] t-aug-Dyall.ae3z
[b]s-aug-Dyall.ae4z
[*]R, CCSD(T)

***Table S2: Static and dynamic polarizabilities (in a.u.) of Group-18 atoms***

| Atom | Freq. (a.u.) | SFX2C1e-LRCC[a] | SFX2C1e-LRCC[b] | Ref. |
|---|---|---|---|---|
| He | 0.072 | 1.388 | 1.380 | 1.383759±0.000013[6] |
| Ne | 0.072 | 2.705 | 2.617 | 2.66110±0.00003[7] |
| Ar | 0.072 | 11.333 | 11.137 | 11.070±0.007 [8] |
| Kr | 0.072 | 17.115 | 16.984 | 16.782±0.005[9] |
| Xe | 0.072 | 27.85 | 27.57 | 27.078±0.050[9] |
| Rn | 0.000 | | 33.09 | 35.04±1.8[10]* |

[a]t-aug-Dyall.ae3z
[b]s-aug-Dyall.ae4z
*R, Dirac, CCSD(T)

**Programmable expressions for Linear Response CCSD in spin-adapted formulation:**

Einstein summation notation is used.

**(a) First ordered perturbed *t* amplitude equations:**

$$(D_i^a)t_i^{a(1)} = A_{ai} - \omega t_i^{a(1)} + t_i^{e(1)}\bar{H}_{ae} - t_m^{a(1)}\bar{H}_{mi} + 2t_m^{e(1)}\bar{H}_{maei} - t_m^{e(1)}\bar{H}_{maie} + \tau_{mi}^{ea(1)}\bar{H}_{me} + \tau_{im}^{ef(1)}\bar{H}_{amef} - \tau_{mn}^{ae(1)}\bar{H}_{mnie}$$

$$(D_{ij}^{ab})t_{ij}^{ab(1)} = A_{abij} - \frac{1}{2}\omega t_{ij}^{ab(1)} + t_i^{e(1)}\bar{H}_{abej} - t_m^{a(1)}\bar{H}_{mbij} + t_{ij}^{eb(1)}\bar{H}_{ae} - t_{mj}^{ab(1)}\bar{H}_{mi} + \frac{1}{2}t_{mn}^{ab(1)}\bar{H}_{mnij} + \frac{1}{2}t_{ij}^{ef(1)}\bar{H}_{abef} + \tau_{mi}^{ea(1)}\bar{H}_{mbej} - t_{mi}^{ea(1)}\bar{H}_{mbje} - t_{im}^{eb(1)}\bar{H}_{maje} + Z_{mi}t_{mj}^{ab} + Z_{ae}t_{ij}^{eb}$$

where

$$\tau_{ij}^{ab(1)} = 2t_{ij}^{ab(1)} - t_{ji}^{ab(1)}$$

$$D_i^a = \bar{H}_{ii} - \bar{H}_{aa} + \omega$$
$$D_{ij}^{ab} = \bar{H}_{ii} + \bar{H}_{jj} - \bar{H}_{aa} - \bar{H}_{bb} + \omega$$

**(b) First ordered perturbed $\lambda$ amplitude equations:**

$$D_i^a\lambda_i^{a(1)} = \eta_i^{a(1)} + \omega\lambda_i^{a(1)} + \lambda_i^{e(1)}\bar{H}_{ea} - \lambda_m^{a(1)}\bar{H}_{im} + \lambda_m^{e(1)}\left(2\bar{H}_{ieam} - \bar{H}_{iema}\right) + \lambda_{im}^{ef(1)}\bar{H}_{efam} - \lambda_{mn}^{ae(1)}\bar{H}_{iemn} - G_{ef}^{7}\left(2\bar{H}_{eifa} - \bar{H}_{eiaf}\right) - G_{mn}^{7}(2\bar{H}_{mina} - \bar{H}_{imna})$$

$$D_{ij}^{ab}\lambda_{ij}^{ab(1)} = P_+(ij/ab)\begin{bmatrix} \eta_{ij}^{ab(1)} + \frac{1}{2}\omega\lambda_{ij}^{ab(1)} + 2\lambda_i^{a(1)}\bar{H}_{jb} - \lambda_j^{a(1)}\bar{H}_{ib} + \lambda_{ij}^{eb(1)}\bar{H}_{ea} - \lambda_{mj}^{ab(1)}\bar{H}_{im} + \frac{1}{2}\lambda_{mn}^{ab(1)}\bar{H}_{ijmn} \\ + \frac{1}{2}\lambda_{ij}^{ef(1)}\bar{H}_{efab} + \lambda_i^{e(1)}\left(2\bar{H}_{ejab} - \bar{H}_{ejba}\right) - \lambda_m^{b(1)}\left(2\bar{H}_{jima} - \bar{H}_{ijma}\right) \\ + 2\lambda_{mj}^{eb(1)}\bar{H}_{ieam} - \lambda_{mi}^{eb(1)}\bar{H}_{jeam} - \lambda_{mj}^{eb(1)}\bar{H}_{iema} - \lambda_{mi}^{be(1)}\bar{H}_{jema} + G_{ae}^{9}\bar{H}_{ijeb} - G_{mi}^{9}\bar{H}_{mjab} \end{bmatrix}$$

where

$$
\begin{aligned}
\eta_{ij}^{ab(1)} &= 2\lambda_i^a A_{jb} - \lambda_j^a A_{ib} + \lambda_{ij}^{eb} A_{ea} - \lambda_{mj}^{ab} A_{im} - t_m^{e(1)} \lambda_j^a L_{mieb} - t_m^{e(1)} \lambda_m^b L_{ijae} \\
&- t_m^{e(1)} \lambda_i^e L_{jmba} + 2t_m^{e(1)} \lambda_j^b L_{imae} - t_m^{e(1)} \lambda_{ij}^{eb} \bar{H}_{ma} - t_m^{e(1)} \lambda_{jm}^{ba} \bar{H}_{ie} - t_m^{e(1)} \lambda_{ij}^{ef} \bar{H}_{fmba} \\
&- t_m^{e(1)} \lambda_{im}^{bf} \bar{H}_{fjea} - t_m^{e(1)} \lambda_{jm}^{fa} \bar{H}_{ejfa} + t_m^{e(1)} \lambda_{ij}^{fb} (2\bar{H}_{fmae} - \bar{H}_{fmae}) + t_m^{e(1)} \lambda_{jm}^{bf} (2\bar{H}_{fiea} - \bar{H}_{fiae}) \\
&+ t_m^{e(1)} \lambda_{in}^{eb} \bar{H}_{jmna} + t_m^{e(1)} \lambda_{ni}^{eb} \bar{H}_{mjna} + t_m^{e(1)} \lambda_{nm}^{ba} \bar{H}_{jine} - t_m^{e(1)} \lambda_{nj}^{eb} (2\bar{H}_{mina} - \bar{H}_{imna}) \\
&- t_m^{e(1)} \lambda_{nj}^{eb} (2\bar{H}_{imne} - \bar{H}_{mine}) + \frac{1}{2} t_{mn}^{ef(1)} \lambda_{ij}^{ef} (ma|J)(J|nb) + \frac{1}{2} t_{mn}^{ef(1)} \lambda_{mn}^{ba} (if|J)(J|je) \\
&+ \frac{1}{2} t_{mn}^{ef(1)} \lambda_{mi}^{fb} (ja|J)(J|ne) + \frac{1}{2} t_{mn}^{ef(1)} \lambda_{im}^{fb} (na|J)(J|je) - \frac{1}{2} t_{mn}^{ef(1)} \lambda_{mj}^{fb} L_{inae} - G_{in}^1 \lambda_{jn}^{ba} \\
&+ G_{af}^1 \lambda_{ij}^{fb} + G_{be}^2 \lambda_{ij}^{ae} - G_{jm}^2 \lambda_{im}^{ab} + 2\lambda_{nj}^{fb} t_{mn}^{ef(1)} L_{imae} - \lambda_{ni}^{fb} t_{mn}^{ef(1)} L_{mjea}
\end{aligned}
$$

$$
\begin{aligned}
\eta_i^{a(1)} &= 2A_{ia} - \lambda_m^a A_{mi} + \lambda_i^e A_{ea} + \lambda_{im}^{fe} A_{feam} - \lambda_{mn}^{ea} A_{ienm} + 2t_m^{e(1)} L_{imae} \\
&+ t_m^{e(1)} \begin{bmatrix} -\lambda_i^e \bar{H}_{ma} - \lambda_m^a \bar{H}_{ie} + 2\lambda_i^f \bar{H}_{fmae} - \lambda_m^f \bar{H}_{fiae} + 2\lambda_m^f \bar{H}_{fiea} - \lambda_i^f \bar{H}_{fmea} \\ -2\lambda_n^e \bar{H}_{mina} + \lambda_n^e \bar{H}_{imna} - 2\lambda_n^a \bar{H}_{imne} + \lambda_n^a \bar{H}_{mine} \end{bmatrix} \\
&+ \tau_{mn}^{ef(1)} \lambda_n^f L_{imae} - \lambda_n^a G_{ni}^3 + \lambda_i^e G_{ea}^3 \\
&+ t_m^{e(1)} \left[ -\lambda_{ni}^{ef} \bar{H}_{mfna} - \lambda_{in}^{ef} \bar{H}_{mfan} - \lambda_{nm}^{af} \bar{H}_{ifne} - \lambda_{mn}^{af} \bar{H}_{ifen} + \frac{1}{2} \lambda_{on}^{ea} \bar{H}_{imno} + \frac{1}{2} \lambda_{on}^{ea} \bar{H}_{mion} \right] \\
&+ t_n^{b(1)} G_{fb}^4 L_{inaf} + t_m^{e(1)} G_{fa}^4 L_{mief} - t_m^{e(1)} G_{ni}^4 L_{mnea} - t_j^{f(1)} G_{nj}^4 L_{inaf} \\
&- G_{mi}^5 \bar{H}_{ma} + G_{ea}^5 \bar{H}_{ie} - t_{mn}^{ef(1)} \lambda_{im}^{gf} \bar{H}_{gnea} - t_{mn}^{ef(1)} \lambda_{mi}^{fg} \bar{H}_{gnae} - t_{mn}^{ef(1)} \lambda_{mn}^{ga} \bar{H}_{gief} - t_{in}^{ef(1)} \lambda_{mn}^{gf} \left( 2\bar{H}_{giae} - \bar{H}_{giea} \right) \\
&- G_{ga}^6 \left( 2\bar{H}_{giae} - \bar{H}_{giea} \right) + t_{mn}^{ef(1)} \lambda_{oi}^{ef} \bar{H}_{mnoa} + t_{mn}^{ef(1)} \lambda_{mo}^{fa} \bar{H}_{inoe} + t_{mn}^{ef(1)} \lambda_{on}^{ea} \bar{H}_{miof} \\
&- G_{mo}^7 \left( 2\bar{H}_{mioa} - \bar{H}_{imoa} \right) - t_{mn}^{ef(1)} \lambda_{no}^{fa} \left( 2\bar{H}_{imoe} - \bar{H}_{mioe} \right)
\end{aligned}
$$

**(c) Equation for response function:**

$$
\begin{aligned}
\langle\langle \mathcal{A}; \mathcal{B} \rangle\rangle &= -\left[A_{ai}\right]^{\mathcal{A}} \left[\lambda_i^{a(1)}\right]^{\mathcal{B}} - \frac{1}{2} \left[P_+(ij/ab) A_{abij}\right]^{\mathcal{A}} \left[\lambda_{ij}^{ab(1)}\right]^{\mathcal{B}} \\
&- 2\left[A_{ia}\right]^{\mathcal{A}} \left[t_i^{a(1)}\right]^{\mathcal{B}} - \lambda_i^a \left[A_{ac}\right]^{\mathcal{A}} \left[t_i^{c(1)}\right]^{\mathcal{B}} + \lambda_i^a \left[A_{ik}\right]^{\mathcal{A}} \left[t_k^{a(1)}\right]^{\mathcal{B}} - \lambda_i^a \left[A_{jb}\right]^{\mathcal{A}} \left[\tau_{ij}^{ab(1)}\right]^{\mathcal{B}} \\
&- \lambda_{kj}^{bc} \left[A_{bcaj}\right]^{\mathcal{A}} \left[t_k^{a(1)}\right]^{\mathcal{B}} + \lambda_{ij}^{ab} \left[A_{kbij}\right]^{\mathcal{A}} \left[t_k^{a(1)}\right]^{\mathcal{B}} + \lambda_{ij}^{ab} \left[A_{ki}\right]^{\mathcal{A}} \left[t_{kj}^{ab(1)}\right]^{\mathcal{B}} - \lambda_{ij}^{ab} \left[A_{bc}\right]^{\mathcal{A}} \left[t_{ij}^{ac(1)}\right]^{\mathcal{B}}
\end{aligned}
$$

and,

$$A_{mi} = \mu_{mi} + t_i^e \mu_{me}$$

$$A_{ia} = \mu_{ia}$$

$$A_{ai} = \mu_{ai} + \mu_{ae} t_i^e - \mu_{mi} t_m^a + \tau_{mi}^{ea} \mu_{me} - t_i^e t_m^a \mu_{me}$$

$$A_{ae} = \mu_{ae} - t_m^a \mu_{me}$$

$$A_{mbij} = t_{ij}^{eb} \mu_{me}$$

$$A_{abei} = -t_{mi}^{ab} \mu_{me}$$

$$A_{abij} = t_{ij}^{eb} A_{ae} - t_{mj}^{ab} A_{mi}$$

$$Z_{ae} = t_m^{f(1)} \left( 2\bar{H}_{amef} - \bar{H}_{amfe} \right) - t_{mn}^{af(1)} L_{mnef}$$

$$Z_{ae} = t_n^{e(1)} \left( 2\bar{H}_{mnie} - \bar{H}_{nmie} \right) - t_{in}^{ef(1)} L_{mnef}$$

$$L_{ijab} = 2(ia|J)(J|jb) - (ib|J)(J|ja)$$

$$L_{ijka} = 2(ja|J)(J|ik) - (ia|J)(J|jk)$$

$$L_{aibc} = 2(ic|J)(J|ba) - (ib|J)(J|ca)$$

$$\tilde{\tau}_{ij}^{ab} = t_{ij}^{ab} + t_i^a t_j^b$$

$$\bar{H}_{me} = f_{me} + t_n^f L_{mnef}$$

$$\bar{H}_{mi} = f_{mi} - t_i^e f_{me} + t_n^e L_{mnie} + \tilde{\tau}_{in}^{ef} L_{mnef}$$

$$\bar{H}_{ae} = f_{ae} - t_m^a f_{me} + t_m^f L_{amef} - \tilde{\tau}_{mn}^{fa} L_{mnfe}$$

$$\bar{H}_{mnij} = (mi|J)(J|nj) + t_j^e (ne|J)(J|mi) + t_i^e (me|J)(J|nj) + \tilde{\tau}_{ij}^{ef} L_{menf}$$

$$\bar{H}_{abef} = (ae|J)(J|bf) - t_m^b (mf|J)(J|ea) - t_m^a (me|J)(J|fb) + \tilde{\tau}_{mn}^{ab} L_{menf}$$

$$\bar{H}_{amef} = (mf|J)(J|ea) - t_n^a (ne|J)(J|mf)$$

$$\bar{H}_{mnie} = (ne|J)(J|mi) + t_i^f (mf|J)(J|ne)$$

$$\bar{H}_{mbej} = (jb|J)(J|me) + t_j^f (me|J)(J|bf) - t_n^b (me|J)(J|nj) - \tilde{\tau}_{jn}^{fb} (nf|J)(J|me) + t_{jn}^{fb} L_{nmfe}$$

$$\bar{H}_{mbje} = (mj|J)(J|be) + t_j^f (mf|J)(J|eb) - t_n^b (ne|J)(J|mj) - \tilde{\tau}_{jn}^{fb} (ne|J)(J|mf)$$

$$\bar{H}_{abei} = (ib|J)(J|ea) - f_{me}t_{mi}^{ab} - t_m^f t_{ni}^{ab} L_{mnfe}$$
$$+t_i^f(ae|J)(J|bf) - t_i^f t_m^a(me|J)(J|bf) - t_i^f t_m^b(mf|J)(J|ae)$$
$$+t_i^f t_{mn}^{ab}(me|J)(J|nf) + t_i^f t_m^a t_n^b(me|J)(J|nf) + \tilde{\tau}_{mn}^{ab}(me|J)(J|ni)$$
$$-t_{im}^{fa}(me|J)(J|bf) - t_{im}^{fb}(mf|J)(J|ea) + t_{mi}^{fb}L_{amef} - t_m^b(mi|J)(J|ae) - t_m^a(ib|J)(J|me)$$
$$+t_m^a t_{in}^{fb}(me|J)(J|nf) - t_m^a t_{ni}^{fb} L_{mnef} + t_m^b t_{ni}^{af}(ne|J)(J|mf)$$

$$\bar{H}_{mbij} = (jb|J)(J|mi) + f_{me}t_{ij}^{eb} + t_n^f t_{ij}^{eb} L_{mnef}$$
$$-t_n^b(mi|J)(J|nj) - t_i^e t_n^b(me|J)(J|nj) - t_j^e t_n^b(ne|J)(J|mi) - t_n^b t_{ij}^{ef}(me|J)(J|nf)$$
$$-t_i^e t_j^f t_n^b(me|J)(J|nf) + \tilde{\tau}_{ij}^{ef}(me|J)(J|fb) - t_{in}^{eb}(me|J)(J|nj) - t_{jn}^{eb}(ne|J)(J|mi) + t_{jn}^{be}L_{mnie}$$
$$+t_j^e(mi|J)(J|be) + t_i^e(me|J)(J|jb) - t_i^e t_{jn}^{fb}(me|J)(J|nf) + t_i^e t_{nj}^{fb} L_{mnef} - t_j^e t_{in}^{fb}(mf|J)(J|ne)$$

$$\lambda_i^a = 2\bar{H}_{ia} + \lambda_i^e \bar{H}_{ea} - \lambda_m^a \bar{H}_{im} + 2\lambda_m^e \bar{H}_{ieam} - \lambda_m^e \bar{H}_{iema} + \lambda_{im}^{ef}\bar{H}_{efam} - \lambda_{mn}^{ae}\bar{H}_{iemn}$$
$$-G_{ef}^9\left(2\bar{H}_{eifa} - \bar{H}_{eiaf}\right) - G_{mn}^9\left(2\bar{H}_{mina} - \bar{H}_{imna}\right)$$

$$D_{ij}^{ab}\lambda_{ij}^{ab} = P_+(ij/ab)\begin{bmatrix} L_{ijab} + 2\lambda_i^a \bar{H}_{jb} - \lambda_j^a \bar{H}_{ib} + \lambda_{ij}^{eb}\bar{H}_{ea} - \lambda_{mj}^{ab}\bar{H}_{im} \\ +\frac{1}{2}\lambda_{mn}^{ab}\bar{H}_{ijmn} + \frac{1}{2}\lambda_{ij}^{ef}\bar{H}_{efab} + \lambda_i^e(2\bar{H}_{ejab} - \bar{H}_{ejba}) - \lambda_m^b(2\bar{H}_{jima} - \bar{H}_{ijma}) \\ +2\lambda_{mj}^{eb}\bar{H}_{ieam} - \lambda_{mi}^{eb}\bar{H}_{jeam} - \lambda_{mj}^{eb}\bar{H}_{iema} - \lambda_{mi}^{be}\bar{H}_{jema} + G_{ae}^{11}L_{ijeb} - G_{mi}^{11}L_{mjab} \end{bmatrix}$$

$$t_i^a = -2t_i^c f_{ck} t_k^a + F_{ac}t_i^c - F_{ki}t_k^a + f_{ia} + 2F_{kc}t_{ki}^{ca} - F_{kc}t_{ik}^{ca}$$
$$+2t_k^c(kc|J)(J|ia) - t_k^c(ki|J)(J|ac) + F_{kc}t_i^c t_k^a + 2t_{ik}^{cd}(kd|J)(J|ac)$$
$$-t_{ik}^{cd}(kc|J)(J|ad) + 2t_i^c t_k^d(kd|J)(J|ac) - t_i^c t_k^d(kc|J)(J|ad)$$
$$-2t_{kl}^{ac}(lc|J)(J|ki) + t_{kl}^{ac}(kc|J)(J|li) - 2t_l^c t_k^a(lc|J)(J|ki) + t_l^c t_k^a(kc|J)(J|li)$$

$$t_{ij}^{ab} = -P_+(ij/ab)\left[t_k^a t_j^c(ki|J)(J|bc) + t_k^b\left[t_j^c(kc|J)(J|ia) + (ak|J)(J|ij)\right]\right]$$
$$+(ia|J)(J|jb) + \tilde{\tau}_{kl}^{ab}W_{klij} + \tilde{\tau}_{ij}^{cd}W_{abcd}$$
$$+P_+(ij/ab)\left[t_{ij}^{cb}L_{ac} - t_{kj}^{ab}L_{ki} + t_{kj}^{cb}(2W_{akic} - W_{akci}) - t_{kj}^{bc}W_{akic} - t_{kj}^{ac}W_{bkci}\right]$$

where

$$F_{ki} = f_{ki} + 2(kc|J)(J|ld)\tilde{\tau}_{il}^{cd} - (kd|J)(J|lc)\tilde{\tau}_{il}^{cd}$$

$$F_{ac} = f_{ac} - 2(kc|J)(J|ld)\tilde{\tau}_{kl}^{ad} + (kd|J)(J|lc)\tilde{\tau}_{kl}^{ad}$$

$$F_{kc} = f_{kc} + \left[2(kc|J)(J|ld) - (kd|J)(J|lc)\right]t_l^d$$

$$W_{klij} = (ki|J)(J|lj) + (lc|J)(J|ki)t_j^c + (kc|J)(J|lj)t_i^c + (kc|J)(J|ld)t_{ij}^{cd} + (kc|J)(J|ld)t_i^c t_j^d$$

$$W_{abcd} = (ac|J)(J|bd) - (kd|J)(J|ac)t_k^b - (kc|J)(J|bd)t_k^a$$

$$W_{akic} = (kc|J)(J|ia) + (kc|J)(J|ad)t_i^d - (kc|J)(J|li)t_l^a$$
$$-\frac{1}{2}(ld|J)(J|kc)t_{il}^{da} - \frac{1}{2}(lc|J)(J|kd)t_{il}^{ad} - (ld|J)(J|kc)t_i^d t_l^a + (ld|J)(J|kc)t_{il}^{ad}$$

$$W_{akci} = (ki|J)(J|ac) + (kd|J)(J|ac)t_i^d - (lc|J)(J|ki)t_l^a - \frac{1}{2}(lc|J)(J|kd)t_{il}^{da} - (lc|J)(J|kd)t_i^d t_l^a$$

$$P_+(ij/ab)\,f(i,j,a,b) = f(i,j,a,b) + f(j,i,b,a)$$

*Where J is the index used to represent the auxiliary dimension.*

- ***Cartesian coordinates (in Å) of the molecules considered in this manuscript:***

**Hydrogen Fluoride (HF):**

```
H      0.00000    0.00000     0.00000
F      0.00000    0.00000     0.91680
```

**Hydrogen Chloride (HCl):**

```
H      0.00000    0.00000     0.00000
Cl     0.00000    0.00000     1.27450
```

**Hydrogen Bromide (HBr):**

```
H      0.00000    0.00000     0.00000
Br     0.00000    0.00000     1.41440
```

**Hydrogen Iodide (HI):**

```
H      0.00000    0.00000     0.00000
I      0.00000    0.00000     1.60916
```

**Iodine monochloride (ICl):**

```
I      0.00000    0.00000     0.00000
Cl     0.00000    0.00000     2.32087
```

**Iodine dimer ($I_2$):**

```
I      0.00000    0.00000     0.00000
I      0.00000    0.00000     2.66630
```

**Na-Li:**

```
Na     0.00000    0.00000     0.00000
Li     0.00000    0.00000     2.81000
```

**K-Cs:**

```
K      0.00000    0.00000     0.00000
Cs     0.00000    0.00000     4.28500
```

**Gold Hydride (AuH):**

```
H     0.00000     0.00000     0.00000
Au    0.00000     0.00000     1.52850
```

**Mercury(II) Chloride ($HgCl_2$):**

```
Hg    0.00000     0.00000     0.00000
Cl    0.00000     0.00000     2.28000
Cl    0.00000     0.00000    -2.28000
```

**Osmium Tetraoxide ($OsO_4$):**

| | | | |
|---|---|---|---|
| O | 0.00000 | 0.00000 | 0.00000 |
| Os | 1.71400 | 0.00000 | 0.00000 |
| O | 2.28534 | 1.61597 | 0.00000 |
| O | 2.28533 | -1.39947 | -0.80800 |
| O | 2.28533 | 1.39948 | -0.80798 |

**Urenium Hexafluoride ($UF_6$):**

| | | | |
|---|---|---|---|
| U | 0.00000 | 0.00000 | 0.00000 |
| F | 0.00000 | 0.00000 | 1.99900 |
| F | 0.00000 | 0.00000 | -1.99900 |
| F | 0.00000 | 1.99900 | 0.00000 |
| F | 0.00000 | -1.99900 | 0.00000 |
| F | 1.99900 | 0.00000 | 0.00000 |
| F | -1.99900 | 0.00000 | 0.00000 |